\documentclass[aps,prb,floatfix,showpacs,twocolumn]{revtex4}
\usepackage{dcolumn,graphicx}
\begin{document}

\title{ Fine structure and size dependence of exciton and bi-exciton
  optical spectra in CdSe nanocrystals }

\author{Marek Korkusinski}
\affiliation{Quantum Theory Group, Institute for Microstructural
  Sciences, National Research Council, Ottawa, Canada, K1A0R6}

\author{Oleksandr Voznyy}
\affiliation{Quantum Theory Group, Institute for Microstructural
  Sciences, National Research Council, Ottawa, Canada, K1A0R6}

\author{Pawel Hawrylak}
\affiliation{Quantum Theory Group, Institute for Microstructural
  Sciences, National Research Council, Ottawa, Canada, K1A0R6}

\begin{abstract}
Theory of electronic and optical properties of exciton and bi-exciton 
complexes confined in CdSe spherical nanocrystals is presented.  
The electron and hole states are computed using atomistic $sp^3d^5s^*$
tight binding Hamiltonian including an effective crystal field
splitting, spin-orbit interactions, and model surface passivation. 
The optically excited states are expanded in electron-hole
configurations and the many-body spectrum is computed in the
configuration-interaction approach.
Results demonstrate that the low-energy electron spectrum is organized
in shells ($s$, $p$, \dots), 
whilst the valence hole spectrum is composed of four
low-lying, doubly degenerate states separated from the rest by a gap.  
As a result, the bi-exciton and exciton spectrum is composed of a
manifold of closely lying states, resulting in a fine structure of
exciton and bi-exciton spectra.  
The quasi-degenerate nature of the hole spectrum results in a correlated
bi-exciton state, which makes it slowly convergent with basis size. 
We carry out a systematic study of the exciton and bi-exciton emission
spectra as a function of the nanocrystal diameter and find that the
interplay of repulsion between constituent excitons and correlation
effects results in a change of the sign of bi-exciton binding energy
from negative to positive at a critical nanocrystal size.
\end{abstract}


\maketitle

\section{Introduction}
Semiconductor nanocrystals (NCs)
(Refs.~\onlinecite{klimov_arpc_2007,nair_geyer_prb08,gur_fromer_science05,nirmal_dabbousi_nature96,brus_nanolett2010,ekimov_hache_josa93,ellingson_beard_nanolett05,scholes_afm08,scholes_rumbles_natmat06,yu_wang_science03,pandey_guyot_jcp2007,gomez_califano_pccp06})
are nano-sized crystalline particles with numbers of atoms of the
order of $10^2$-$10^5$.
NCs with controlled and tunable sizes as well as good optical
properties are fabricated in a colloidal growth
process.\cite{yin_alivisatos_nature05} 
This makes them excellent candidates for use in low-cost
optoelectronic applications, including solar cells,
biomarkers,\cite{fu_gu_con05,alivisatos_natbio_04,chan_nie_science98} 
light emitting
diodes,\cite{colvin_schlam_nature94,dabbousi_bawendi_apl95,chanyawadee_lgoudakis_advmat09}
photodetectors,\cite{konstantatos_sargent_natnano2010,sukhovatkin_hinds_science09}
single-photon sources in
quantum cryptography,\cite{pattantyus_qiao_nanolett09} or
lasers.\cite{klimov_mikhaelovsky_science00,weisbuch_jcg94}
In particular, it has been recently demonstrated that the optical gain
in NCs can be blocked, created, and tuned by engineering the NC
confinement\cite{klimov_ivanov_nature07} or the type of multiexciton
complex active in the stimulated emission
process.\cite{cooney_sewall_prl09}  

The NCs are considered as a promising material for
the optically active media in solar cells.
They offer a potential way to increase the efficiency of solar cells
due to their tunable parameters amenable to
optimization,\cite{hillhouse_beard_cocis09,wadia_alivisatos_est09,luther_law_nanolett08,choi_lim_nanolett09,johnston_pattantyus_apl08}
as well as by generation  of multi-exciton complexes (MEG)
following absorption of a single high-energy
photon.\cite{nozik_physE_2002,rabani_baer_nanolett08,schaller_pietryga_nanolett07,franceschetti_an_nanolett06,nair_bawendi_prb07,ellingson_beard_nanolett05,scholes_rumbles_natmat06,nozik_arpc01}
During MEG a high-energy photon with energy 
of at least twice the semiconductor bandgap, $2E_g$, is absorbed
creating an excited state, which can be described as a
superposition of configurations with one and more
electron-hole pairs.\cite{nozik_physE_2002}
Alternatively, we can think of exciting a single exciton, which is
then converted via Coulomb interactions into additional interacting
electron-hole pairs. 
Energy relaxation of these multi-exciton complexes results in multiple
carriers at the bottom of the conduction and the top of the valence
bands.  
These multi-exciton states decay into exciton states by Auger
processes, limiting the number of additional charges generated in the
MEG process. 
The process of conversion of a single exciton into multiple
electron-hole pairs competes with the phonon-assisted relaxation of
exciton energy.\cite{nozik_arpc01}   
Since the original report by Schaller and
Klimov,\cite{schaller_klimov_prl2004}  the MEG process
has been reported in  PbSe, PbS, PbTe, CdSe, InAs, and Si 
NCs,\cite{nozik_cpl2008} with efficiency reaching 700\% (seven
electron-hole pairs out of one photon).\cite{schaller_sykora_nanolett2006}
However, proper assessment of the MEG efficiency in these experiments
is nontrivial.\cite{nair_bawendi_prb07,pandey_guyot_jcp2007,mcguire_sykora_nanoletters2010}   
The potential explanation of MEG has been given by Shabaev, Efros, and
Nozik\cite{shabaev_efros_nanolett06} and alternative interpretation
proposed by Zunger and co-workers\cite{franceschetti_an_nanolett06}
and others.\cite{mcguire_joo_acr08,silvestri_agranovich_prb10,schaller_agranovich_natphys2005}

The lowest-energy MEG process involves conversion of an
excited exciton into a low-energy bi-exciton following
absorption of a photon with energy of $\sim2E_g$.
Therefore, a detailed study of the electronic and
optical properties of the bi-exciton is needed.
To date, theoretical attention has been focused mainly on the
properties of low-energy exciton states, with the CdSe NCs being the
most studied system.
The electronic and optical properties of an exciton (X) confined in a NC
have been explored utilizing the multi-band $k\cdot p$
method,\cite{ekimov_hache_josa93,efros_rosen_prb96},
tight-binding\cite{leung_pokrant_prb98,chen_whaley_prb04,sapra_sarma_prb04,delerue_allan_prb05}
and empirical pseudopotential methods.\cite{wang_zunger_jpchem98,franceschetti_fu_prb99,califano_franceschetti_prb07} 
These studies show a fine structure in the low-energy states of X
originating from the electron-hole exchange, with the energy gap
between the lowest - dark, and the higher - bright states of the order
of several meV.

Identification of bi-exciton (XX) signatures in emission spectra is
complicated by the presence of the inhomogeneous broadening in the
ensemble measurements on NCs.
However, one can measure consistently the XX binding energies both in
ensemble measurements in
CdSe\cite{sewall_franceschetti_prb09,achermann_hollingsworth_prb03,caruge_chan_prb04,bonati_mohamed_prb05}
and CdS NCs.\cite{klimov_ivanov_nature07}
First single-NC experiments have also been
reported.\cite{osovsky_cheskis_prl09} 
In particular, the exciton fine structure has been studied in
individual NCs as a function of the magnetic
field.\cite{htoon_crooker_prl2009,furis_htoon_prb2006} 
Thus, the state of experimental techniques is approaching that in
epitaxially grown quantum dots, for which single-dot experiments, revealing
details of the fine structure of multiexciton complexes, are
now a standard.\cite{michler-book}
One of experimental tools utilized to obtain spectroscopic information
about bi-excitons confined in NCs, the transient absorption, involves
measuring with a short probe pulse the change of absorption induced by
the pump laser pulse.\cite{klimov_arpc_2007,sewall_franceschetti_prb09}
By utilizing the transient absorption technique one can probe states
of XX via both emissive and absorptive experiments, which opens a
possibility of probing the fine structure of XX directly.\cite{sewall_franceschetti_prb09,sewall_cooney_prb06,sewall_cooney_jcp08}
The experimental results of Ref.~\onlinecite{sewall_franceschetti_prb09}
obtained on CdSe dots with the diameter of $5.6$ nm have been compared 
to the results of empirical pseudopotential calculations carried out
on dots with diameter of $3.8$ and $4.6$ nm.
However, to our knowledge, no systematic study of the
dependence of the bi-exciton spectra on the parameters of the CdSe NCs
have been carried out.\cite{shumway_franceschetti_prb01} 
The quantitative analysis of these systems is computationally
challenging due to the NC size.
With $\sim 10^5$ electrons, NCs are too large for {\em ab initio}
methods.
On the other hand, the $k\cdot p$ methods are not accurate enough to
capture important atomistic details, such as the asymmetry of the
crystal lattice or the surface effects.
This necessitates the use of the semi-empirical atomistic methods in
the theoretical analysis.

Here we utilize the atomistic tight-binding approach to perform a
systematic study of the electronic and optical properties of an X and
XX confined in a single CdSe NC as a function of NC size.
We illustrate our calculations on a spherical NC with the
diameter of $3.8$ nm for comparison with the empirical pseudopotential
of Ref.~\onlinecite{sewall_franceschetti_prb09}.
To this end we utilize the QNANO computational
platform.\cite{korkusinski_zielinski_jap09}  
The atomistic single particle states are used in computation of
the Coulomb matrix elements, describing the carrier-carrier
interactions, and the optical dipole elements.
The many-body multi-exciton states are computed using exact
diagonalization techniques.
 
The results show the $s$ and $p$ shells in the low-energy electron
spectrum  as expected from a single-band effective mass theory.
For holes we find  a complex spectrum, consisting of a band of four
Kramers doublets forming a quasi-degenerate hole shell separated
from the remaining hole levels by a gap. 
The energy separation of these states is much smaller than the
characteristic Coulomb hole-hole interaction matrix elements.
Therefore we predict the bi-exciton (XX) spectrum to be composed of a
manifold of closely lying correlated states of two electrons
residing mainly on the $s$-shell and a correlated complex of two holes
occupying almost degenerate hole states, resulting in a
fine structure of bi-exciton optical spectra.   
The exciton (X) spectrum, on the other hand, reveals the fine
structure determined both by the hole shell degeneracy and the
electron-hole exchange interaction. 
We find that the correlated character of both the X and
XX systems makes the computations challenging, with large basis sizes
necessary to obtain converged values of their energies.
In this work we discuss how this fine structure influences the
absorption and emission spectra of both X and XX complexes.
We find that for small NCs (with diameter below $4$ nm) the bi-exciton
is unbound, whilst for larger NCs it is bound.
Also, the order of X and XX emission peaks with model inhomogeneous
broadening depends on temperature. 
We show that due to details of the electronic structure and assignment
of oscillator strengths, the thermal population of excited XX states
leads to a shift of the inhomogeneously broadened XX peak to lower
energies, whilst the analogous process leads to the increase of the X
emission energy.
The shifts are of the order of tens of meV and may lead to the
reversal of the order of emission peaks.

\section{Model}

We analyze the electronic and optical properties of electrons and
holes confined in a single, spherical CdSe nanocrystal.
The calculations are carried out utilizing the QNANO computational
platform and consist of the following steps: (i) the definition of
the geometry and composition of the nanostructure on the atomistic
level, (ii) the calculation of single-particle quasi-electron and
quasi-hole states using the 20-band $sp^3d^5s^*$ tight-binding (TB)
model,  
(iii) the computation of many-body energies and states of $N$
quasi-electron-quasi-hole pairs in the configuration-interaction (CI)
approach (in this case $N=1$ and 2), and (iv) calculation of
emission and absorption spectra using Fermi's Golden Rule. 
A detailed review of the QNANO package and the computational procedure
is given in Ref.~\onlinecite{korkusinski_zielinski_jap09}.

\subsection{Atomistic tight-binding description 
  of a nanocrystal}

The computational procedure starts with a definition of the positions
of all atoms present in the system.
The underlying crystal lattice of the CdSe nanocrystal is taken to be in
wurtzite modification, which is built out of two 
hexagonal closely packed (hcp) sublattices, one made up of cations and
another of anions, shifted with respect to one another.
As a result, each atom is surrounded by four nearest neighbors.
The hcp structure is described by two lattice parameters, $a$ and $c$,
which, in principle, are independent.
In this work we assume however that the nearest neighbors of each atom
form a perfect tetrahedron.
This relates the two lattice parameters with one another such that
we have $a = \sqrt{3\over8} c$.
With $c = 0.70109$ nm (Ref. \onlinecite{landolt}), this gives
$a = 0.42933$ nm, as compared to the experimental value of $0.42999$
nm.
If one parametrizes all the distances with the lattice constant $c$,
the positions of the four atoms in the wurtzite unit cell are as
follows:
we have two anions, in our case Selenium, 
at $(0,0,0)$ and $(\sqrt{6}/8,\sqrt{2}/8,1/2)c$, and two cations,
in our case Cadmium, at $(\sqrt{6}/8,\sqrt{2}/8,1/8)c$ and
$(0,0,5/8)c$. 

Calculation of the single-particle states is carried out in the 
linear combination of atomic orbitals (LCAO)
approximation, in which the carrier wave function is written as a
linear combination 
\begin{equation}
\Psi _{i} \left(\vec{r}\right)=\sum _{R=1}^{N_{AT} }
\sum _{\alpha=1}^{20}A_{R\alpha }^{(i)} \varphi_{\alpha}
\left(\vec{r}-\vec{R}\right)  
\label{LCAO}
\end{equation}
of atomistic orbitals of type $\alpha$ localized on the atom
$R$, with $N_{AT}$ being the total number of atoms in the
system. 
In our $sp^3d^5s^*$ model we deal with 10 doubly spin-degenerate basis
orbitals on each atom.   
The coefficients $A_{R\alpha }^{(i)}$ determining the $i$th
single-particle state as well as the corresponding single-particle
energies are found by diagonalizing the semi-empirical atomistic
TB Hamiltonian
\begin{eqnarray}
H_{TB} &=& \sum _{R=1}^{N_{AT} }\sum _{\alpha =1}^{20}\varepsilon
_{R\alpha } c_{R\alpha }^{+} c_{R\alpha }   +\sum _{R=1}^{N_{AT} }\sum
_{\alpha =1}^{20}\sum _{\alpha '=1}^{20}\lambda _{R\alpha \alpha '}
c_{R\alpha }^{+} c_{R\alpha '}   \nonumber \\
&+&\sum _{R=1}^{N_{AT} }\sum
_{R'=1}^{nn}\sum _{\alpha =1}^{20}\sum _{\alpha '=1}^{20}t_{R\alpha
  ,R'\alpha '} c_{R\alpha }^{+} c_{R'\alpha '}  ,   
\end{eqnarray}
in which the operator $c_{R\alpha }^{+}$ ($c_{R\alpha }$) creates
(annihilates) the particle on the orbital $\alpha $ of atom
$R$. 
The Hamiltonian is parametrized by the on-site orbital
energies $\varepsilon _{R\alpha } $, spin-orbit coupling constants
$\lambda _{R\alpha \alpha '} $, and hopping matrix elements
$t_{R\alpha ,R\alpha '}$ connecting different orbitals located at
neighboring atoms. 
In our model we use the nearest
neighbor approximation, and therefore do not capture directly the
crystal field splitting, which is due to the symmetry breaking on the
level of third nearest neighbors. Following
Ref.~\onlinecite{leung_pokrant_prb98} we include the crystal field
splitting  in an approximate manner by detuning the energy of the
orbital $p_z$ from that of the orbitals $p_x$, $p_y$ which remain
degenerate.  
The TB parameters are obtained by calculating the band
structure of bulk CdSe and fitting the band edges and effective masses
at high symmetry points of the Brillouin zone to the values obtained
experimentally or by {\em ab initio} calculations.
The parametrization used in this work is given in Table~\ref{tbparameters}.
\begin{table}
\caption{\label{tbparameters} Tight-binding parameters for CdSe used
  in this work. All values are in eV, and the notation follows that of
Slater and Koster}
\begin{ruledtabular}
\begin{tabular}{ld}
Parameter & \text{Value} \\
$E_s^{a}$ & -10.9438 \\
$E_{px,py}^{a}$ & 1.3131 \\
$E_{pz}^{a}$ &  1.2795 \\
$E_{d}^{a}$ & 6.9721 \\
$E_{s*}^{a}$ &  7.5610 \\
$\lambda_{SO}^{a}$ & 0.1307 \\ 
$E_s^{c}$ &  0.7855 \\
$E_{px,py}^{c}$ &  4.7247 \\
$E_{pz}^{c}$ &  4.6844 \\
$E_{d}^{c}$ &  6.4424 \\
$E_{s*}^{c}$ &  6.4704 \\
$\lambda_{SO}^{c}$ &  0.1568 \\
$V_{ss}$ & -0.9470 \\
$V_{sa,pc}$ & 2.6220 \\
$V_{pa,sc}$ & 1.8608 \\
$V_{pp\sigma}$ & 3.1287 \\
$V_{pp\pi}$ & -0.5674 \\
$V_{sa,s*c}$ & -0.0001 \\
$V_{s*a,sc}$ & -0.1685 \\
$V_{pa,s*c}$ & 0.4694 \\
$V_{s*a,pc}$ & 0.0004 \\
$V_{s*,s*}$ & -0.0937 \\
$V_{sa,dc}$ & -0.0649 \\
$V_{da,sc}$ & -0.0079 \\
$V_{pa,dc\sigma}$ & -0.0137 \\
$V_{da,pc\sigma}$ & -0.0005 \\
$V_{pa,dc\pi}$ & 0.0053 \\
$V_{da,pc\pi}$ & 0.0004 \\
$V_{s*a,dc}$ & -0.0748 \\
$V_{da,s*c}$ & -0.0121 \\
$V_{dd\sigma}$ & -0.0007 \\
$V_{dd\pi}$ & 0.1479 \\
$V_{dd\delta}$ & -0.1834 \\
\end{tabular}
\end{ruledtabular}
\end{table}
Using this parametrization we obtain the following parameters of the
bulk band structure.
The bandgap $E_g = 1.83$ eV (we fit to the low-temperature data), the
crystal field splitting 
$E_{CFS}=0.0254$ eV and the spin-orbit splitting $\Delta_{SO}=0.444$
eV correspond closely to the experimental values of $E_g=1.83$ eV,
$E_{CFS}=0.026$ eV and $\Delta_{SO}=0.429$ eV
(Ref.~\onlinecite{landolt}).
The electron effective masses are $m_e^*(M)=0.133 m_0$ 
towards the $M$
point, and $m_e^*(A)=0.134 m_0$ towards the $A$ point, 
while the measured value for both directions is $m_e^*=0.13 m_0$.
In the highest valence subband, the effective mass towards the $M$
point is $m_{h1}^*(M)=0.455 m_0$, 
while the measured value is $0.45 m_0$.
In the same subband the mass towards the $A$ point is 
$m_{h1}^*(A)=1.443 m_0$ while the measured value is $1.17 m_0$.
Finally, in the second valence subband we compute the effective mass
towards the $M$ point to be $m_{h2}^*(M)=0.851 m_0$ which is close to
the experimental value of $0.9 m_0$.

Figure~\ref{fig1} shows the band structure of CdSe bulk computed with
the above TB parameters (a) compared to the band structure
obtained in DFT calculation (b), in which the conduction band was rigidly
shifted by $1.562$ eV to reproduce the experimental value of the gap.
Note that in our parametrization the on-site energies of $d$ orbitals
on both the anion and the cation are above the energies of orbitals
$s$ and $p$.
This is in contrast to several other parametrizations accounting for the $d$
orbitals,\cite{sapra_shanthi_prb02,visvanatha_sapra_prb05,gurel_akinci_tsf08}
where the cation $d$ orbitals lie below the
$s$ and $p$ orbitals.
In such parametrizations it is possible to reproduce the flat $d$-band
visible in Fig.~\ref{fig1}(b) at the energy of about $-8$ eV.
Since all our $d$ orbitals lie high in energy, in our bulk band
structure in Fig.~\ref{fig1}(a) the $d$-band is not present.
Our choice of the placement of $d$ orbitals was dictated by 
the fact that, according to the GW calculations, the admixture of the
low-lying $d$ orbitals in the wave functions corresponding to the top
of the valence band in the $\Gamma$ point is
negligible.\cite{zakharov_rubio_prb94}
Since we set out to study the properties of several lowest exciton and
bi-exciton states, we concentrate on an accurate reproduction of band
edges rather than deeper bands.
Moreover, the small size of our NCs necessitates an accurate
description of the conduction band across the Brillouin zone, which in
turn entails the use of the high-energy $s*$ orbitals.
These orbitals are taken to have higher on-site energies than the
respective high-energy $d$ orbitals on both the cation and anion.
We thus have to account for all these high-lying orbitals and neglect
the low-energy cation $d$-band  in order to treat both
types of atoms on equal footing.

Figure~\ref{fig:bulk_dos} shows the bulk density of states (DOS)
computed using three methods: 
the result of the density functional (DFT) LCAO calculation using
the SIESTA package\cite{siesta_webpage} (top panel),
the plane-wave approach used in the
package ``Exciting''\cite{exciting_webpage} (middle panel), 
 and the DOS resulting from our TB approach
(bottom panel).
All three panels shows the DOS within the energy range of two gap
energies into the valence and conduction bands, i.e., the range of
energies of interest for the multiexciton generation process.
Thus, our TB model gives results consistent with the two other, 
{\em ab initio} approaches up to $3$ eV into each of the conduction
and valence bands.

The TB Hamiltonian in the above parametrization is used to compute the
single-particle states in a spherical nanocrystal.
The positions of all atoms in such a system are determined by cutting
a spherical sample out of a bulk semiconductor, without any surface
relaxation effects.
The dangling bonds on the surface of the nanocrystal are passivated by
the procedure involving the following steps:
(i) rotation from the
$s-p_x-p_y-p_z$ basis to that of $sp^3$ hybridized orbitals, (ii)
identification of the directions of resulting bonds and application of
an energy shift of $25$ eV to those that are unsaturated, and (iii)
inverse rotation into the $s-p_x-p_y-p_z$ basis.\cite{lee_oyafuso_prb03}

\subsection{Description of interacting electrons and holes confined in the nanocrystal}

The excited states of the NC are expanded in electron and
hole pair configurations. 
The electrons are defined as occupied states in the conduction band
and holes as empty states in the valence band. 
With the operator $c_{i}^{+} $ ($c_{i} $) creating (annihilating) an
electron on the single-particle state $i$, while the operator
$h_{\alpha }^{+} $ ($h_{\alpha} $) creating (annihilating) a hole on the
single-particle state $\alpha $, the excited states $|\nu\rangle$ are
written as  
\begin{equation}
|\nu\rangle
=\sum _{i,\alpha} B_{i,\alpha}^{\nu} 
c_{i}^{+} h_{\alpha }^{+} {\left| 0 \right\rangle}
+\sum _{i,j,\alpha,\beta} C_{i,j,\alpha,\beta}^{\nu} c_{i}^{+} c_{j}^{+}
h_{\alpha }^{+} h_{\beta }^{+} {\left| 0 \right\rangle} + \dots,
\end{equation}
where ${\left| 0 \right\rangle}$ is the ground state of the NC.
The amount of mixing among the configurations with different number of
excitations is defined by the amplitudes 
$B_{i,\alpha}^{\nu}$, $C_{i,j,\alpha,\beta}^{\nu}$
and depends on the energy of the state.
The ground exciton state, whose energy is of order of the
semiconductor gap $E_g$, will be built predominantly out of single pair
excitations, with a negligible contribution from the two-pair
(energy at least of order of $2E_g$) or higher configurations.
On the other hand, the two-pair excitations may be mixed with highly
excited single-pair configurations with similar energies.
In this work we shall treat the number of quasi-particles as a good
quantum number when labeling the states of the NC.
A detailed analysis of the mixing effects will be presented elsewhere.

The Hamiltonian of interacting $N_e$ electrons and $N_h$
holes distributed on the single-particle states is
\begin{eqnarray}
H &=& \sum _{i}\varepsilon_{i} c_{i}^{+} c_{i}
    +\sum _{\alpha }\varepsilon _{\alpha } h_{\alpha }^{+} h_{\alpha }
    +\frac{1}{2} \sum _{ijkl}{\left\langle ij \right|} V_{ee} {\left|
        kl \right\rangle} c_{i}^{+} c_{j}^{+} c_{k} c_{l} \nonumber\\
    &+&\frac{1}{2} \sum _{\alpha \beta \gamma \delta }{\left\langle
        \alpha \beta  \right|} V_{hh} {\left| \gamma \delta
      \right\rangle} h_{\alpha }^{+} h_{\beta }^{+} h_{\gamma }
    h_{\delta }  \nonumber \\
&-&\sum _{il}\sum _{\beta \gamma }\left({\left\langle i\beta  \right|}
      V_{eh} {\left| \gamma l \right\rangle} -{\left\langle i\beta
        \right|} V_{eh} {\left| l\gamma  \right\rangle}
    \right)c_{i}^{+} h_{\beta }^{+} h_{\gamma } c_{l} .
\label{interacting_hamiltonian}
\end{eqnarray}
In Eq.~(\ref{interacting_hamiltonian}) the first two terms account for the
single-particle energies, the third and fourth terms describe the
electron-electron and hole-hole Coulomb interactions, respectively,
and the last term introduces the electron-hole direct and exchange
interactions.  
The Coulomb matrix elements are computed
using the single-particle TB wave functions. 
In these computations we separate (i) the on-site terms arising from
the scattered particles residing on the same atom, 
(ii) the nearest-neighbor (NN) terms involving orbitals localized on
adjacent atoms, and 
(iii) the long-distance terms describing scattering between more
remote atoms.
Using the general form of our LCAO wave functions (\ref{LCAO}), each of
these three elements can be written as follows:
\begin{eqnarray}
{\left\langle ij \right|} V_{ee} {\left| kl \right\rangle}
&=& V_{ons} + V_{NN} + V_{long}, 
\label{coulomb_elements}\\
V_{ons} &=& \sum_{R=1}^{N_{AT}}{
\sum_{\alpha\beta\gamma\delta=1}^{20}{
A_{R\alpha}^{(i)*}
A_{R\beta}^{(j)*}A_{R\gamma}^{(k)}
A_{R\delta}^{(l)} }} \nonumber\\
&\times& 
\langle\left. R\alpha,R\beta \right|
{e^2\over\epsilon_{ons}|\vec{r}_1-\vec{r}_2|}
\left| R\gamma, R\delta\right.\rangle, \\
V_{NN} &=& \sum_{R_i=1}^{N_{AT}}{\sum_{R_j}^{NN}{
\sum_{\alpha\beta\gamma\delta=1}^{20}{
 A_{R_i\alpha}^{(i)*}
 A_{R_j\beta}^{(j)*} A_{R_j\gamma}^{(k)}
A_{R_i\delta}^{(l)}}}} \nonumber\\
&\times& 
\langle\left. R_i\alpha,R_j\beta \right|
{e^2\over\epsilon_{NN}|\vec{r}_1-\vec{r}_2|}
\left| R_j\gamma, R_i\delta\right.\rangle, 
\label{coulomb_nn_slater} \\
V_{long} &=& \sum_{R_i=1}^{N_{AT}}{\sum_{R_j}^{remote}{
\sum_{\alpha\beta=1}^{20}{
 A_{R_i\alpha}^{(i)*}
 A_{R_j\beta}^{(j)*} A_{R_j\beta}^{(k)}
A_{R_i\alpha}^{(l)}}}} \nonumber\\
&\times& 
{e^2\over\epsilon_{long}|\vec{R}_i-\vec{R}_j|}. 
\label{coulomb_remote}
\end{eqnarray}
and analogously for the hole-hole and electron-hole interactions.
The necessary integrals in the on-site and nearest-neighbor terms are
computed by approximating the atomistic functions $|R,\alpha\rangle$
by Slater orbitals.\cite{slater-pr1930}
Note that in the above formulas we have assumed the two-center
approximation.
In an attempt to simulate the distance-dependent dielectric
function,\cite{franceschetti_fu_prb99,wang_califano_prl03,moreels_allan_prb10,ogut_burdick_prl03,delerue_lannoo_prb03}
each of these terms is scaled by a different dielectric constant
$\epsilon$.
Typically we take $\epsilon_{ons}=1$ and $\epsilon_{long}=5.8$, the
latter one being the bulk CdSe dielectric constant, while
$\epsilon_{NN}=2.9$, i.e., half of the bulk
CdSe value.
In what follows we shall present two computations, one with the
nearest neighbor term assuming the form identical to the remote term,
and another one, with the nearest-neighbor term as in
Eq.~(\ref{coulomb_nn_slater}).

\subsection{Calculation of optical spectra}

Once the many-body states of the system of interacting electron-hole
pairs are established, we calculate the emission spectra utilizing the
Fermi's Golden Rule 
\begin{equation}
I\left(\omega \right)=\sum _{f,i}P_i(T)\left|{\left\langle f \right|}
  P_{X} {\left| i \right\rangle} \right|^{2} \delta \left(E_{i}
  -E_{f} -\hbar \omega \right) ,
\label{fgoldrule}
\end{equation}
where $E_i$ is the energy of the initial state of  $N$
excitons, $E_f$ is that of the final state of  $N-1$
excitons, and the sum is carried over all possible final states. 
The temperature-dependent factor $P_i$ describes thermal population of
levels of the initial exciton complex.
The transition intensity is determined by the interband polarization
operator, which for the polarization $x$ is defined as 
$P_{X} =\sum _{i j}d_{i j }^{(x)} c_{i} h_{j }  $. 
The single particle dipole elements $d_{i j }^{(x)} $ are defined 
as $d_{i j}^{(x)} =\int d\vec{r}\phi
_{h,j}^{*} \left(\vec{r}\right)x\phi _{e,i} \left(\vec{r}\right) $.
The polarization operators for polarizations $y$ and $z$ are
defined analogously.
In our TB approach, the dipole matrix elements can be evaluated in the
form:
\begin{eqnarray}
d_{i j}^{(x)} &=& \sum_{R = 1}^{N_{AT}} \sum_{\alpha=1}^{20}{
A_{R\alpha}^{*(j)}A_{R\alpha}^{(i)} R_x} \nonumber\\
&+& \sum_{R}^{N_{AT}} \sum_{\alpha\beta=1}^{20}{
A_{R\alpha}^{*(j)}A_{R\beta}^{(i)}} 
 \int{ d\vec{r}
\phi^{*}_{\alpha}\left( \vec{r}   \right)
x \phi_{\beta}\left(\vec{r}\right)}.
\end{eqnarray}
The integrals involving orbitals from the nearest and further
neighbors are neglected.
Absorption spectra are obtained using a formula analogous to
Eq.~(\ref{fgoldrule}), only the polarization operator $P_{X}$ is
replaced by its Hermitian conjugate.
In this case the initial state describes the system of $N$ excitons,
with the appropriate thermal occupation of levels,
while the final state contains $N+1$ electron-hole pairs.

\section{Single-particle states in the spherical nanocrystal}

Figure~\ref{fig2} illustrates the single-particle properties of a
spherical CdSe nanocrystal of diameter of $3.8$ nm, whose atomistic image
is shown in the inset of Fig.~\ref{fig2}(c). 
The system consists of $1028$ atoms, with the Cd (Se) atoms rendered in
blue (red). 
The energies of the single-particle electron and hole states obtained
in the tight-binding calculation are shown in Figs.~\ref{fig2}(a) and
\ref{fig2}(b), respectively. 
The structure of electron states is typical for a
spherical quantum confinement: the ground state of the $s$
symmetry is separated by a large gap (about $270$ meV) from three
states of the $p$ symmetry. For the valence
holes, however, we find {\em four} closely lying states,
highlighted in Fig.~\ref{fig2}(b) by the blue rectangle, separated from
the remainder of the spectrum by a gap of about $120$ meV. 
This structure of the hole states is due to the interplay of the
spin-orbit interaction and the crystal field splitting.
The characteristic gap is robust and appears also for NCs with larger
diameters, as illustrated in Fig. \ref{fig2}(c). The existence of four
closely-lying hole states appears to be in agreement with results of earlier 
empirical pseudopotential calculations.\cite{franceschetti_fu_prb99,sewall_franceschetti_prb09}

In the case of electron states, whose energies are shown in
Fig.~\ref{fig2}(a), the $p$ shell consists of three levels:
almost degenerate $p_x$ and $p_y$ states at a higher energy, and a
single non-degenerate $p_z$ level at a slightly lower energy.
This is a signature of the wurtzite symmetry of the NC,
which differentiates between the $+z$ and $-z$ directions,
leading to a corresponding asymmetry in the electron
wave function.

Insight into the symmetry of the four hole states
emphasized in Fig.~\ref{fig2}(b) can be gained by
computing the dipole matrix elements $d_{ij}^{(x)}$
built out of the $i$-th electron and $j$-th hole
states, with $y$ and $z$ matrix elements constructed analogously. 
In Fig.~\ref{fig4}(a) we plot the joint optical density
of states (JDOS), i.e., magnitude of dipole elements 
$|d_{ij}|^2$ versus the energy
gap between the ground electron ($i=1$) and the four lowest
hole states. 
For polarizations $x$ and $y$ we obtain
four nonzero elements, while for polarization $z$ the JDOS
consists of only two peaks. 
This structure of JDOS can be understood
by approximating the atomistic wave functions as products of the
envelope and Bloch part, as is done in the $k\cdot p$ model. 
 Since the envelope
function changes slowly on interatomic distances, one typically
approximates the dipole element by a product of the overlap of
electron and hole envelope functions and an integral involving the
Bloch components and the position operator appropriate for the
$x$, $y$, or $z$ polarization. 
The electron
ground state is built out of $s$-type atomistic orbitals
modulated by an $s$-type envelope, while the hole states are
built out of $p$-type atomistic orbitals.
Due to the spin-orbit mixing the envelope functions of the hole
are mixtures of different symmetries.\cite{rego_hawrylak_prb97}
However, this projectional analysis will extract the part of the
envelope function of the hole which is of the same symmetry as the
electron envelope function (in this case, symmetry $s$).

Using this approximation let us first analyze the lowest (H1) and
highest (H4) JDOS maxima. 
They are present in the $x$ and $y$
polarizations, but absent in the $z$ polarization.
This means that the $s$-like term in the hole envelope function
is associated with the Bloch functions consisting of $p_x$ and $p_y$,
but not $p_z$ atomic orbitals.
The overlap of the electron and hole envelope functions is
large for H1, but very small for H4, which indicates that the 
 $s$-like component dominates in the envelope function of the hole
 ground state, while the state H4 is of a different symmetry. 
The two middle JDOS peaks, H2 and H3, appear in
all polarizations, indicating that the Bloch components of the
corresponding hole functions are combinations of all three atomistic
$p$ orbitals.
Of those two, H2 is consistently stronger than H3, which suggests 
that the hole state H2 has a larger, and H3 - a smaller
$s$-like component. 

Further confirmation of this assignment of symmetries
is obtained by computing the dipole matrix elements between the
$p$ electron states and the four hole states. 
This procedure probes the $p$-like component in the hole envelope
functions.
The elements
are shown in Fig.~\ref{fig4}(b) as red, blue, and green bars, with the
assignment of colors explained in Fig.~\ref{fig4}(c). 
We find that the ground state H1 gives a negligible dipole
matrix element with either of the three electron $p$ states, which
confirms that the state H1 is of the $s$ type. 
On the other hand, the state H4 presents large
dipole elements, which indicates that it has a dominating $p$-like
component in its envelope function.
 
As previously with the $s$-type electron state, the two middle peaks,
H2 and H3, appear consistently in both polarizations, but now H3 is
larger. 
We conclude that the hole states H2 and H3 are mixtures of $s$ and
$p$-type envelopes. 
This assignment of symmetries is only approximate, as the
details of the underlying crystal lattice and surface roughness break
the rotational symmetry of the nanocrystal. 
However, a visual inspection of charge densities of the hole states
suggests a picture consistent with the above analysis.

Note that all single-particle states are Kramers doublets, whose
degeneracy is due to the time-reversal symmetry of the single-particle
Hamiltonian. 
In what follows we shall distinguish the states forming the doublet by
arrows up and down, respectively. 
Due to the spin-orbit interaction these labels cannot be identified
with particle spins, but rather with Bloch total angular momenta. 

\section{Excitonic complexes in the nanocrystal}
\subsection{Exciton}

In order to find the energies and states of the exciton (X) we generate
all possible electron-hole configurations in the single-particle
basis, write the full Hamiltonian (\ref{interacting_hamiltonian}) in a
matrix form in the basis of these configurations, and diagonalize this
matrix numerically.
Construction of the Hamiltonian requires knowledge of the Coulomb
electron-hole scattering matrix elements.
Typically one distinguishes two types of Coulomb matrix elements: the
``direct'' and the ``exchange'' ones, the latter originating from the
antisymmetric character of the many-body wave function.
This distinction is particularly clear in the case of diagonal matrix
elements, i.e., those arising when one computes the expectation value
of the Coulomb operator against any configuration.
In this case the direct terms can involve pairs of particles with
different spin, while the exchange elements connect particles with the
same spins.
Due to the spin-orbit interaction present in our TB model the
single-particle states cannot be characterized by a definite spin.
Moreover, as already mentioned, all single-particle states are in
reality Kramers doublets, and any linear combination of the two
constituent states is also a good eigenstate of the TB Hamiltonian.
In order to be able to separate and analyze the Coulomb elements, we
perform a rotation of each pair of states forming the Kramers doublet
so as to optimize the expectation value of the Pauli $\sigma_z$
operator.
With the states thus prepared we compute the Coulomb elements using
the formulas (\ref{coulomb_elements}). 

Let us now comment on the magnitudes of various Coulomb matrix
elements for our NC with diameter of $3.8$ nm.
We will discuss these elements in two cases, depending on the treatment
of the nearest-neighbor contributions: (i) the case when these
contributions are computed exactly using Slater orbitals, as shown
in formula (\ref{coulomb_nn_slater}), (ii) the case when they are
expressed simply by the formula (\ref{coulomb_remote}) as for remote
centers.
In each case we scale the nearest neighbor contribution by the
dielectric constant of $2.9$, i.e., half of the CdSe bulk value.

If we denote the states composing the lowest-energy electron doublet
as $|1_e\downarrow\rangle$ and $|1_e\uparrow\rangle$, and the
analogous pair of hole 
states as $|1_h\downarrow\rangle$ and $|1_h\uparrow\rangle$,
we find the direct elements:
$\langle 1_e\downarrow 1_h\downarrow | V_{eh} | 1_h\downarrow 1_e\downarrow \rangle = 
\langle 1_e\downarrow 1_h\uparrow | V_{eh} | 1_h\uparrow 1_e\downarrow \rangle = 212.76$ meV in the case (i), 
and $220.52$ meV in the case (ii).
These elements define the interaction energy of an electron-hole
configuration $c^+_{1\downarrow}h^+_{1\downarrow}|0\rangle$ 
and $c^+_{1\downarrow}h^+_{1\uparrow}|0\rangle$,
respectively, where $|0\rangle$ denotes quasi-particle vacuum.
The ``spin-flip'' electron scattering, described, e.g., by an element
$\langle 1_e\downarrow 1_h\downarrow | V_{eh} | 1_h\downarrow 1_e\uparrow \rangle$, is not possible, as
the value of this element is negligibly small.
However, due to the much stronger spin mixing of the hole states
resulting from the spin-orbit interaction one might expect
that the transitions involving the hole spin flip should be possible.
In fact, here the only elements of note are
$\langle 1_e\downarrow 1_h\downarrow | V_{eh} | 4_h\uparrow 1_e\downarrow \rangle = 
\langle 1_e\downarrow 1_h\uparrow | V_{eh} | 4_h\downarrow 1_e\downarrow \rangle = 0.26$ meV in the case (i), 
and $0.30$ meV in the case (ii).

The scattering among the hole states, with the electron staying on the
same level and without hole spin flip,
$\langle 1_e\downarrow 1_h\downarrow | V_{eh} | 2_h\downarrow 1_e\downarrow \rangle$ and 
$\langle 1_e\downarrow 1_h\downarrow | V_{eh} | 3_h\downarrow 1_e\downarrow \rangle$
is very small.
However, the scattering onto the fourth hole Kramers doublet,
$\langle 1_e\downarrow 1_h\downarrow | V_{eh} | 4_h\downarrow 1_e\downarrow \rangle$,
is much larger and its absolute value amounts to $5.2$ meV in the case
(i) and $5.8$ meV in the case (ii).
Such an element describes the Coulomb coupling between configurations
$c^+_{1\downarrow}h^+_{1\downarrow}|0\rangle$ and $c^+_{1\downarrow}h^+_{4\downarrow}|0\rangle$.
Also, scattering with hole transfer between the second and third
Kramers doublet is sizeable and amounts to about $8.02$ meV in the
case (i) and $8.86$ meV in the case (ii).
The energy scales set by these Coulomb elements are to be compared
with the energy separation of the hole states, which ranges from about
$5$ to about $15$ meV. 
Thus, the diagonal Coulomb electron-hole terms, i.e., those
that do not lead to a change of the electron-hole configuration, are 
about $20$ times larger than the separation of the hole states.
On the other hand, the scattering elements, describing the change of
configuration, are approximately of the same order as this separation.
Therefore at this point it is not clear whether the ground state of
the X can be approximated by a single configuration or it is
rather a correlated system, with the hole spread out among the four
lowest Kramers doublets. 

The second type of the Coulomb matrix elements in play is the
electron-hole exchange.
Let us specify the exchange elements involving the lowest Kramers
doublets - one for the electron and one for the hole.
As these elements are spin sensitive, let us first point out that the
TB model allows us to compute only the states of electrons in the
conduction and valence bands.
In order to describe the optics of our system in the usual language of
quasi-particles, we have to change the treatment of the valence band by
renaming the missing valence electron ``spin up'' into the valence hole ``spin
down''.
Having this in mind, and working for the moment in the language of
electrons only, we find that the absolute value of the element
$\langle 1_e\downarrow 1_h\downarrow | V_{eh} | 1_e\downarrow 1_h\downarrow \rangle =
\langle 1_e\uparrow 1_h\uparrow | V_{eh} | 1_e\uparrow 1_h\uparrow \rangle = 6.36$ meV, while
$\langle 1_e\downarrow 1_h\uparrow | V_{eh} | 1_e\downarrow 1_h\uparrow \rangle =
\langle 1_e\uparrow 1_h\downarrow | V_{eh} | 1_e\uparrow 1_h\downarrow \rangle = 0.01$ meV
in the case (i).
In the case (ii) the absolute value of these elements is $7.2$ meV
and $0.01$ meV, respectively.
In the language of quasi-particles these diagonal exchange terms
describe the interaction correction to the electron-hole pairs with
opposite spin (the former two elements) and parallel spin (the latter two
elements).
Since the electron-hole exchange interaction enters the total
Hamiltonian with the positive sign, the quasi-particle pairs with
antiparallel spins (i.e., the optically active ones) will have higher
energy than those with parallel spins. 
The remaining exchange terms within this manifold of states are
offdiagonal, and their absolute values do not exceed $1$ $\mu$eV.
As a result, due to the electron-hole exchange we expect a fine
structure of the X composed of two pairs of states, separated by
a gap of several meV, and each pair nearly degenerate.

To conclude the discussion of the electron-hole Coulomb matrix
elements, we comment on how these elements are impacted by 
the difference in treatments of nearest neighbors, i.e., the cases (i)
and (ii).
We find that the treatment (ii) gives consistently larger magnitudes
of the elements, however the difference is only about 5\% in the
direct terms and about 20\% in the smallest exchange terms.
From the general formulas for Coulomb elements given by 
Eqs.~(\ref{coulomb_nn_slater}) and (\ref{coulomb_remote})
we see that the nearest-neighbor term scales as the number of atoms
$N_{AT}$, while the remote term scales as $N_{AT}^2$.
As a result, the change in treatment of nearest neighbors will be more
visible in smaller nanocrystals.
In what follows we will employ the simplified treatment (ii), as the
resulting change of energies is small compared to other energy scales
in the system, and the simplification of the treatment of nearest
neighbors leads to a considerable speedup in calculations.

Let us now move on to constructing the correlated states of the
interacting electron-hole pair.
As was already mentioned, we accomplish this by diagonalizing the
Hamiltonian (\ref{interacting_hamiltonian}) set up in the basis of
electron-hole configurations.
With $1028$ atoms present in the system, and the TB basis of $20$
orbitals per atom, we can distribute our particles on $20560$
single-particle states, out of which we have $3700$ hole and 
$16860$ electron states.
As a result, there exist $62.3\times 10^6$ excitonic configurations.
Since in this work we focus on the low-energy excitonic configurations
only, instead of dealing with the full basis we shall build the
electron-hole configurations out of single-particle states closest to
the bandgap.
The computational effort grows rapidly with the increase of the number
of single-particle basis involved in the calculation.
The most time- and resource-intensive part is the computation of
Coulomb matrix elements, as each element involves $\sim N_{AT}^4$
operations, and for $M_e$ electron and $M_h$ hole states we require
$\sim M_e^4$ electron-electron elements,
$\sim M_h^4$ hole-hole elements, and
$\sim M_e^2 M_h^2$ electron-hole elements.

The evolution of the X spectra as a function of the basis size
is visualized in Fig.~\ref{fig:x-baseladders}(a).
In the left-hand panel we show the X energies resulting from the
diagonalization of the Hamiltonian built using $M_e=2$ electron states
(i.e., the lowest, s-shell Kramers doublet) and $M_h=8$ hole states
(i.e., the lowest four Kramers doublets separated from the rest
of the hole spectrum by a gap).
This results in $16$ electron-hole configurations.
In the middle panel we include more hole states by increasing $M_h$ to
$28$, whilst in the right-hand panel we compute with $M_e=8$ (i.e.,
the s and p shells) and $M_h=28$.
This increase of the single-particle basis gives respectively $56$ and
$224$ configurations.
We see, overall, that as the basis is increased, the energy of the
lowest level decreases, but not by a large amount compared to the
bandwidth of the excitonic states.
Moreover, the excitonic states are grouped into blocks separated by
gaps.
The lowest energy block, consistent throughout the three spectra, is
built out of configurations from the lowest electronic doublet and the
four lowest hole doublets of single-particle states.
The second block, apparent in the middle panel, involves the hole
residing on higher single-particle states, and the gap separating it
from the lower section is consistent with the gap in the
single-particle hole spectrum.
Finally, in the right-hand panel we see two spectra from the middle
panel, stacked on top of one another.
Further, the top half of this ladder of states is denser than the
bottom half.
Such a distinct structure of the spectra is due to a large gap between
the electron single-particle s and p shells.
As a result, the third block from the bottom is composed of the hole
residing on the four lowest single-particle levels, but the electron
occupying one of the six p-shell levels. 
Similarly, the highest block contains configurations with an electron
on the $p$ shell and the hole on states deeper in the valence band.

In Fig.~\ref{fig:x-baseladders}(b) we show the dependence of the X
ground-state energy on the hole basis size for three cases: electron
on the $s$ shell only (black squares), on the $s$ and $p$ shells (red
circles), and on the $s$, $p$, and $d$ shells (green triangles).
We see that as a function of $1/M_h$ the X energy follows a
power law and it is not completely converged even for the largest hole
basis used ($M_{h,max}=128$ states).
We notice also a marked decrease of the X energy as subsequent
electronic shells enter the picture.
As the electron spreads to the p shell, the energy drops by about $12$
meV, while allowing the electron to spread to the d shell results in a
smaller decrease.
Note that the overall drop in energy is about $20$ meV, most of it
accomplished by using the three electronic shells, $s$, $p$, and $d$
(altogether 18 states) and increasing the
hole basis to $M_h=28$ states.
The energy drop is of the order of Coulomb electron-hole direct
scattering matrix elements and is somewhat larger than the separation
of the lowest four hole single-particle states, but it is an order of
magnitude smaller than the diagonal direct Coulomb elements describing
the electron-hole attraction.

In view of the slow convergence of X energies it is necessary to
establish a scheme for extraction of their converged values.
To this end we first extrapolate each of the curves from
Fig.~\ref{fig:x-baseladders}(b) to zero (i.e., infinite number of hole
basis states). 
Then we use the energies obtained for each $M_e$ to extrapolate to
infinite number of electron states.
In our case, the extrapolated ground-state energy of the exciton is
$E_X^{\infty}=2.100$ eV.

Because of the commensurability of the energy change due to
correlations and the single-particle energy scale we now examine the
spectral content of the several low-lying X states
in the case of $M_e=8$ and $M_h=28$, i.e., in the regime when the
energy is already relatively close to convergence.
Figure~\ref{fig:X-converged-content}(a) shows the lowest section of the
energy spectrum of such an X, while panels (b) 
illustrate schematically the configurations dominant in the respective
wave functions.
We find that the two lowest X states are nearly degenerate (to
within $0.1$ $\mu$eV) and are composed predominantly (in 95\%) of the
configurations with the electron and hole occupying their respective
lowest single-particle states, $1_e$ and $1_h$, and assuming the same
spin character (i.e., approximately ``spin parallel'').
Such configurations can be written as $c^+_{1\downarrow}h^+_{1\downarrow}|0\rangle$ and
$c^+_{1\uparrow}h^+_{1\uparrow}|0\rangle$ and
are shown schematically in the
upper left-hand and upper right-hand panel of 
Fig.~\ref{fig:X-converged-content}(b), respectively
The next pair of states is found $7.3$ meV higher in energy, this gap
being due predominantly to the electron-hole exchange.
This pair is degenerate to within $1$ $\mu$eV.
The constituent states are composed of configurations, in which the
electron and the hole have ``opposite spins''.
One of such configurations,
$c^+_{1\downarrow}h^+_{1\uparrow}|0\rangle$ is shown in the lower
left-hand panel of Fig.~\ref{fig:X-converged-content}(b).
The other one, $c^+_{1\uparrow} h^+_{1\downarrow}|0\rangle$, is 
shown in the lower right-hand panel of that Figure.

\subsection{Bi-exciton}

We now proceed to calculating the energies and wave functions of a
system of two electron-hole pairs forming a bi-exciton (XX).
Since now we deal with pairs of carriers of the same type, we need to
establish the electron-electron and hole-hole matrix elements.
In computations we also consider the two cases of treating the
nearest-neighbor contributions, as discussed previously for the
electron-hole elements.
Let us consider the electron-electron elements first.
The diagonal element defining the interaction energy of the
two-electron configuration on the lowest single-particle levels
$c^+_{1\uparrow}c^+_{1\downarrow}|0\rangle$ is
$\langle 1_e\downarrow, 1_e\uparrow |V|1_e\uparrow, 1_e\downarrow \rangle = 197.79$ meV in the case (i)
and $203.71$ meV in the case (ii).
It is somewhat smaller (by about 5\%) than the fundamental
electron-hole element discussed in the previous Section.
A similar element for the holes, defining the interaction energy of
the hole pair $h^+_{1\uparrow}h^+_{1\downarrow}|0\rangle$ is
$\langle 1_h\downarrow, 1_h\uparrow |V|1_h\uparrow, 1_h\downarrow \rangle = 271.77$ meV in the case (i)
and $271.78$ meV in the case (ii).
It is much larger than the fundamental electron-electron and
electron-hole elements.
We find  this to be the case for all sizes of spherical CdSe
nanocrystals studied (from $2$ nm to $7$ nm).
A possible reason for this disparity between various types of matrix
elements lies in the difference of charge densities corresponding to
the electron and hole single-particle ground states.
Figure~\ref{fig3}(a) shows the vertical cross-section of the
ground-state charge density for the electron computed by the QNANO
package. 
As can be seen, this density is distributed across the entire crystal,
it is largest in the center and tapers off towards the surfaces.
Small irregularities in this image are due to the lack of the symmetry
plane of the nanocrystal, which is built out of $11$ layers of atoms.
This is why we find a finite density on the lowest atomic layer, while
the top layer appears to carry no charge.
An analogous profile for the ground hole state is shown in
Fig.~\ref{fig3}(b).
Here we see a clear lack of symmetry, with the maximum charge located
in the lower half of the nanocrystal.
Also, the hole appears to occupy a much smaller volume than the
electron does.
Note that neither of the wave functions is centered in the NC.
This is due to wurtzite structure of NC, particularly due to polarity of 
the (0001) direction which results in the development of internal 
dipole moment. 
We have confirmed the existence of such a dipole in our DFT
calculations (not shown here), however we find that the direction and
strength of this dipole is sensitive to the type of ligands used to
passivate the NC surface.
The decreased spatial extent of the hole state leads to a large
magnitude of the Coulomb repulsion of two holes placed on the lowest
Kramers doublet, since the charge density of these two states is
identical.
In the electron case the more uniform spread of the density across the
crystal diminishes the electron-electron element.
The electron-hole element is also decreased, as we deal here with a
relatively localized hole interacting with a distributed electron charge.

Let us now look at the scattering matrix elements, which describe the
Coulomb coupling between different configurations.
For holes the largest element transferring the particle from the
lowest Kramers doublet is
$\langle 1_h\downarrow, 1_h\uparrow |V|1_h\uparrow, 4_h\downarrow \rangle = 2.73$ meV in the case (i)
and $2.62$ meV in the case (ii).
To assess the strength of this element in a meaningful way, let us
first analyze briefly a two-hole configuration $h_{4\uparrow}^+h_{1\downarrow}^+|0\rangle$.
Compared to the fundamental configuration $h^+_{1\uparrow}h^+_{1\downarrow}|0\rangle$, the
excited configuration is created by moving the hole from 
the first to the fourth Kramers doublet
and so its single-particle energy is higher by about $34$ meV.
On the other hand, the interaction energy of the excited two-hole
configuration is given by the matrix element
$\langle 1_h\downarrow, 4_h\uparrow |V|4_h\uparrow, 1_h\downarrow \rangle$, which is $220.26$ meV
in the case (i) and $220.73$ meV in the case (ii).
That is, in this excited configuration the holes repel by about $50$
meV weaker than in the fundamental one.
So, altogether, the configuration $h_{4\uparrow}^+h_{1\downarrow}^+|0\rangle$ is {\em lower
  in energy,} even though it has a larger single-particle energy part.
The energy difference between the two configurations is then only
about $17$ meV, suggesting that the offdiagonal scattering matrix
elements will lead to the appearance of strongly correlated hole-hole
states.

Let us now account for the presence of the two electrons.
The electron-hole attraction gives a negative contribution to the
total energy of the system.
In any XX configuration we have four constituent terms, as each
electron attracts two holes. 
For this discussion we need also the electron-hole matrix element
$\langle 1_e\downarrow, 4_h\uparrow |V|4_h\uparrow, 1_e\downarrow \rangle = 195.38$ meV in case (i) and
$201.39$ meV in case (ii).
Comparing this element with the fundamental electron-hole direct
element given in the previous section we see that the electron
attracts the hole on the level $4$ by about $18$ meV weaker than
it does the hole on the ground single-particle level.
Now we are in a position to compare the energies of configurations
$c^+_{1\uparrow}c^+_{1\downarrow}h^+_{1\uparrow}h^+_{1\downarrow}|0\rangle$ and
$c^+_{1\uparrow}c^+_{1\downarrow}h^+_{4\uparrow}h^+_{1\downarrow}|0\rangle$.
Owing to the electron-hole direct terms we find that the former, i.e.,
the fundamental configuration of the two electron-hole pairs, is about
$55$ meV lower in total energy than that with one excited hole.
As we see, the final alignment of levels results from cancellations of
large interaction terms of comparable magnitude, and as such will be
sensitive to the details of many-body computation.

The smallest contribution to scattering comes from the
electron-electron interaction.
Transfer of one electron from the $s$-shell orbitals to the $p$-shell
orbitals due to this interaction is described, e.g., by an element
$\langle 1_e\downarrow, 1_e\uparrow |V|1_e\uparrow, 2_e\downarrow \rangle = 0.44$ meV in the case (i)
and $0.49$ meV in the case (ii).
This is to be compared with the energy gap between the $s$ and $p$
electronic shells, which amounts to $270$ meV.
So, the spread of electrons onto the $p$ shell due to the
electron-electron interaction is expected to be small.

We now proceed to diagonalizing the two-pair Hamiltonian as a function
of the size of the single-particle basis.
Figure~\ref{fig:XX-baseladders}(a) shows the energy levels of the system
with $M_e=2$, $M_h=8$ (left), $M_e=2$, $M_h=28$ (middle), and
$M_e=8$, $M_h=28$ (right).
In the first case we populate only the lowest electronic Kramers
doublet and the four lowest hole Kramers doublets.
As a result we can create $28$ configurations. 
The left-hand panel of Fig.~\ref{fig:XX-baseladders}(a) shows all
resulting XX eigenenergies.
As we increase the hole basis, and later on also the electron basis,
we allow one or both particles of each type to populate higher-energy
single-particle states.
This results in a buildup of the density of XX states at higher
energies, which is clearly visible in the middle and right-hand panels
of Fig.~\ref{fig:XX-baseladders}(a).
Also, the low-lying XX energy states appear to shift down in energy
by tens of meV.
To analyze this shift in greater detail, in 
Fig.~\ref{fig:XX-baseladders}(b) we plot the energies of the
XX states with dominant singlet-singlet configuration
$c^+_{1\uparrow}c^+_{1\downarrow}h^+_{1\uparrow}h^+_{1\downarrow}|0\rangle$
as a function of the size of the hole basis in three cases:
with both electrons on the $s$ state (black squares),
with the electrons allowed to spread onto the $p$ shell (red circles)
and with the electrons populating the $s$, $p$, and $d$ shells
(green triangles).
Note that the state under consideration is not always the ground state
of the system, which reflects the simple analysis outlined above.
This state stabilizes as the ground state for hole basis of at least
$M_h=28$ states, while for smaller $M_h$ excitation of one of the
holes is preferred.

As we can see, the energy decreases steadily if we increase the hole
basis size but keep a constant number of electron states.
On the other hand, for a constant hole basis one large drop takes
place as we increase the electron basis from $2$ to $8$ states.
Upon its further increase to $18$ states the energy change is much
less significant.
A systematic study of the convergence of ground state energy is much
more difficult here, as for the basis $M_e=18$, $M_h=124$ we already
deal with $1.17\times 10^6$ two-pair configurations.
Using the procedure analogous to that described in the previous
Section, we extrapolate the ground-state XX energy to the limit of
infinite basis and obtain $E_{XX}^{\infty}=4.229$ eV.
Figure~\ref{fig:XX-baseladders}(a) shows that in spite of the
substantial energy change of the ground state, 
the block of lowest $28$ states appears to be separated from
the remaining spectra, suggesting that the configurations with lowest
single-particle energy contribute to their respective eigenvectors the
most.
To demonstrate this, in Fig.~\ref{fig:XX-converged-content} we analyze the
spectral content of several lowest eigenstates of the system with
$M_e=8$, $M_h=28$. 
Figure~\ref{fig:XX-converged-content}(a) shows the lowest $28$ energy
levels, consistent with the number of possible configurations created
out of lowest electron and hole single particle blocks.
As we can see, the lowest $27$ states are found just within a $65$ meV
window, i.e., a fraction of the value of the fundamental Coulomb
matrix elements.
This is the central result of this work.
We find that in the case of spherical CdSe nanocrystals the peculiar
arrangement of hole single-particle levels together with Coulomb
interactions lead to the appearance of a fine structure of bi-exciton
levels.

In Fig.~\ref{fig:XX-converged-content}(b) we show configurations
dominant in the lowest four XX states.
The first two states, denoted respectively as (I) and (II), 
correspond to the two configurations analyzed before and
behave as
we predicted previously using the single-configuration arguments.
The configuration predominant in the ground state is the
``singlet-singlet'' one, shown in panel (I), and created by placing
the pairs of carriers on lowest possible single-particle levels.
The first excited state, on the other hand, is based on the
configuration where the hole is excited to the fourth Kramers
doublet, shown in the panel (II).
Note that the gap between these two states, amounting to about $9$
meV, is of the order of the electron-hole exchange energy,
but it originates from correlations rather than exchange.
The first excited state also contains significant admixtures of
configurations, in which the holes occupy the two middle Kramers
doublets in the low-energy section (not shown).
At even higher energy we find a pair of nearly degenerate states, in
which the holes have aligned ``spins''.
The dominant configurations in these states are shown in panels (III)
and (IV), respectively.

\subsection{Optical spectra of exciton and bi-exciton}

Having described the electronic properties of the X and XX confined in
the NC we now proceed to discussing their optical spectra.
We will conduct our analysis for the systems built upon $M_e=8$
electron and $M_h=28$ hole single-particle states.
We start with the absorption spectrum of the exciton, shown in
Fig.~\ref{fig:Xabsorption} and focus on the lowest $16$ spectral lines
of that spectrum, i.e., those corresponding to the electron on the $s$
shell and the hole on one of the four lowest Kramers doublets.
We find that the two lowest, nearly degenerate states of the exciton are
optically dark irrespective of polarization, while the 
second pair of states is bright, but only in polarizations $x$ and
$y$.
The two pairs of states form the fine
structure of the exciton, in which the energy gaps are due to the
electron-hole exchange.
In these two polarizations we also find a third, even larger maximum
at a slightly higher energy.
Analysis of the spectral content of the corresponding X eigenstate
reveals that it consists of a mixture of configurations in which the
hole resides on the second and third Kramers doublet, with a smaller
admixture of the fundamental configurations shown in
Fig.~\ref{fig:X-converged-content}(b)(II).
Each of these configurations contributes constructively to the large
final absorption amplitude of this state.
The spectrum in polarization $z$ is different: it consists only of one
large maximum, larger than any of the maxima in polarizations $x$ and $y$.
This large amplitude comes from the predominant distribution of the
hole on the second single-particle Kramers doublet, which introduces a
large oscillator strength in the $z$ polarization, as evident from
Fig.~\ref{fig4}(a).
The general properties of the exciton spectrum, the multiplicities of the
states and their oscillator strength obtained in our calculation agree
well with previous empirical
pseudopotential,\cite{franceschetti_fu_prb99} 
tight-binding\cite{leung_pokrant_prb98}
and qualitative $k\cdot p$ calculations.\cite{efros_rosen_prb96} 

The absorption spectrum discussed above is equivalent to the emission
spectrum of the X at high temperature, i.e., when the occupations of
all X energy levels are similar.
However, at low temperature only the lowest states will be occupied,
and the emission spectrum will consist predominantly of the lowest
line of the exciton.
As this exciton state is dark, its radiative lifetime is expected to
be very long.
This suggests that owing to the characteristic alignment of the hole 
single-particle levels the X emission spectrum will sensitively depend
on the temperature, with the dominant maxima appearing at higher
energy as the temperature is increased.

Let us move on to computing the optical spectra of the XX.
Figure~\ref{fig:XXoptics} shows its absorption (two top panels)
and emission spectra (third panel from the top). 
Calculation of the absorption spectrum involves preparing the
single exciton system in the bright, second excited state (top panel)
or dark, ground state (second panel from the top) and adding the second
electron-hole pair to form the ground and excited XX states. 
Due to the optical selection rules, the carriers composing the photo-created
electron-hole pair must have antiparallel spins.

Let us first discuss the absorption spectra involving an addition of
the second electron-hole pair to the bright exciton. 
We find that this spectrum is composed of several peaks, one at
energy $2.115$ eV, and denoted as $XX_0$,
the second one at energy about $2.124$ eV, and
the third one at energy about $2.136$ eV. The low-energy peak
corresponds to addition of an optically active electron-hole pair to
the bright exciton configuration as in
Fig.~\ref{fig:X-converged-content}(b)(I) and formation of the ground
state bi-exciton $XX_0$ as in Fig.~\ref{fig:XX-converged-content}(b)(I). 
The two higher-energy peaks correspond to the formation of an excited
bi-exciton, in which the holes are redistributed among the four lowest
Kramers doublets, however retaining their ``spin unpolarized'' character.

If the exciton is prepared in the ground, dark state, the XX absorption
spectrum (Fig.~\ref{fig:XXoptics}, second panel from the top) is
dominated by two 
groups of peaks, one around the energy of about $2.13$ eV, denoted as
$XX^*$,
and the second at energy $2.144$ eV, indicating the formation of
excited XX states. 
This is because at any lower energy the optical selection rules
prevent us from adding a photoexcited, spin-unpolarized electron-hole
pair to the spin-parallel ground X state. 
The resulting XX states contain spin-polarized holes.

In the calculation of the emission spectrum (third panel of
Fig.~\ref{fig:XXoptics}) we prepare the XX in its ground,
singlet-singlet state. 
We find that the emission spectrum is dominated by one maximum,
corresponding to the bright exciton final state [(II) in
Fig.~\ref{fig:X-converged-content}(b)]. 
It is accompanied by a small maximum at  lower energy. 
Note that the main maximum is found at the same energy as the
low-energy  absorption peak of the bright exciton (top panel). 
However, there is an energy gap between the emission and dark exciton
absorption peaks, due to the fine structure of both the exciton and
bi-exciton low-energy spectra.

Finally, the bottom panel of Fig.~\ref{fig:XXoptics} shows the
emission spectra of the exciton. 
The low-energy peak  corresponds to the dark, ground state $X_D$,
whilst the high-energy peak denotes the radiative transition from the
bright, excited state $X_B$ of the exciton.

Since in our calculations we have not accounted for the dynamics of
carriers, and in particular the relaxation processes, the above
emission and absorption spectra should be understood as 
presentation of oscillator strengths for various transitions rather
than candidates for direct comparisons with experimental data.
To make this point clear, in Fig.~\ref{fig:abs_emis_processes}
we present two possible absorptive and emissive scenarios in our
system.
In Figure~\ref{fig:abs_emis_processes}(a) and (b) we assume that the
relaxation from the excited, bright state $X_B$ to the ground, dark 
state $X_D$ is faster than the radiative recombination from either of
these two states.
Panel (a) shows the absorption from the vacuum to the XX state.
We start with the optical creation of an exciton in its bright state
(the red arrow), followed by its relaxation to the dark state (the
black arrow).
According to Fig.~\ref{fig:Xabsorption}, the absorption process should
take place at the energy of about $2.122$ eV.
We can now create the XX, but only in one of its excited states, such
as the one denoted as $XX^*$ in Fig.~\ref{fig:XXoptics} (second panel
from the top). 
This absorption event should be observable at the energy of about
$2.128$ eV, i.e., we predict the XX binding energy in this absorptive
process to be {\em negative}.

The emissive cascade under the assumption of fast relaxation is shown
in Fig.~\ref{fig:abs_emis_processes} (b).
We start with XX in its ground state $XX_0$, recombining radiatively and
leaving the system with the X in its excited, bright state $X_B$.
According to Fig.~\ref{fig:XXoptics} (third panel from the top), this
process is seen in the emission as a maximum at $2.115$ eV.
The exciton further relaxes to the dark state $X_D$ and may recombine
with a very long lifetime.
The recombination energy appears to be also $2.115$ eV, but its
degeneracy with the XX emission energy is here accidental and cannot
be treated as the universal property of the NCs.
To summarize, assuming the fast X relaxation we find the absorptive XX
binding energy to be $-6$ meV, while the emissive binding energy is
zero.

Let us now consider the situation in which the X relaxation from its
bright to dark state is the slowest process.
The absorption processes under this condition are visualized in
Fig.~\ref{fig:abs_emis_processes} (c).
Here we create the X in its bright state $X_B$ at the energy $2.122$
eV, and then immediately afterwards the XX in its ground state  $XX_0$.
According to Fig.~\ref{fig:XXoptics} (top panel), the latter
transition should be seen as an absorption maximum at $2.115$ eV,
producing the absorptive XX binding energy of $+7$ meV.
The emissive cascade under the same condition is shown schematically
in Fig.~\ref{fig:abs_emis_processes} (d).
As the cascade utilizes the same X and XX states as the absorptive
process, we expect the XX emissive binding energy in our system to be
also equal to $+7$ meV. 

Let us now present a detailed comparison of the XX and X emission spectra.
As already mentioned, the exciton recombines to vacuum, and therefore
the position of the respective emission peak equals to the energy of
the X.
In the simplest treatment we approximate the X state by one
configuration $|X_0\rangle=c^+_{1\downarrow} h^+_{1\downarrow}|0\rangle$.
The emission energy of such a state, neglecting for now the
electron-hole exchange, is
$E_{X0}=\varepsilon_{1}^{(e)}+\varepsilon_{1}^{(h)}
-\langle 1_e\downarrow,1_h\downarrow|V_{eh}|1_h\downarrow,1_e\downarrow\rangle$.
The bi-exciton, on the other hand, recombines to the final-state
exciton.
The energy of the fundamental XX configuration
$|XX_0\rangle=c^+_{1\uparrow}c^+_{1\downarrow}h^+_{1\uparrow}h^+_{1\downarrow}|0\rangle$ is
$E_{XX0}=2\varepsilon_1^{(e)}+2\varepsilon_1^{(h)}
+\langle 1_e\downarrow,1_e\uparrow|V_{ee}|1_e\uparrow,1_e\downarrow\rangle
+\langle 1_h\downarrow,2_h\uparrow|V_{hh}|1_h\uparrow,1_h\downarrow\rangle
-4\langle 1_e\downarrow,1_h\downarrow|V_{eh}|1_h\downarrow,1_e\downarrow\rangle$
and accounts for  the repulsion of the like carriers and attraction of
each pair of opposite carriers.
The position of the XX fundamental emission peak can then be evaluated
as
$\Omega_{XX} = E_{XX0}-E_{X0} = \left(
\varepsilon_1^{(e)}+\varepsilon_1^{(h)}-\langle 1_e\downarrow,1_h\downarrow|V_{eh}|1_h\downarrow,1_e\downarrow\rangle
\right)
+ \left( \langle 1_e\downarrow,1_e\uparrow|V_{ee}|1_e\uparrow,1_e\downarrow\rangle - \langle
  1_e\downarrow,1_h\downarrow|V_{eh}|1_h\downarrow,1_e\downarrow\rangle \right )
+ \left( \langle 1_h\downarrow,1_h\uparrow|V_{hh}|1_h\uparrow,1_h\downarrow\rangle- \langle
  1_e\downarrow,1_h\downarrow|V_{eh}|1_h\downarrow,1_e\downarrow\rangle \right )$.
It consists of the exciton energy $E_{X0}$ and self-energy corrections.
If the electron-electron and hole-hole elements are equal to the
electron-hole terms, the self-energy corrections cancel out and the XX
emission peak matches that of the exciton.
However, as demonstrated above, the hole-hole repulsive interaction is
significantly larger than the electron-hole element, and so from that
simple analysis we expect the XX peak to occur at the energy {\em
  higher} than that of X, i.e., we expect the XX to be unbound, as is
found in some epitaxial dots (see e.g. 
Ref.~\onlinecite{reimer_dalacu_jpcs2010}).

Let us now compare the emission peak positions of X and
XX.\cite{achermann_hollingsworth_prb03,klimov_ivanov_nature07}  
Figure~\ref{fig:XandXX-hitemp} shows these spectra on the same energy
scale.
The bars in the two top panels show the complete X emission spectrum
assuming equal occupation of all levels (infinite temperature), while
the black and red curves account for finite temperature effects.
At the temperature of $4$ K (top left panel) only the ground state of
the exciton will be occupied, as it is separated from the excited
states by the electron-hole exchange gap.
As a result, the low-temperature X recombination  is forbidden
optically, resulting in a long excitonic lifetime.
Assuming long integration  time, and accounting for a model
inhomogeneous broadening, such a recombination maximum is
schematically shown in Fig.~\ref{fig:XandXX-hitemp}(top left panel)
with a black line.
Note that in this graph only the position of this peak is meaningful,
since, as we have already mentioned,  we do not model dynamical
phenomena taking place in the system. 
On the other hand, as the temperature is increased to $300$K, excited
X states are populated, resulting in optically allowed excitonic
recombination. 

The high-temperature emission maximum computed assuming the
Maxwell-Boltzmann thermal distribution of carriers and 
including model inhomogeneous broadening, is shown with the red
continuous line in the top right panel of Fig.~\ref{fig:XandXX-hitemp}.
As already discussed, this maximum is shifted towards higher energies 
by about $20$ meV.

Let us now move on to the XX spectra (the two bottom panels of
Fig.~\ref{fig:XandXX-hitemp}). 
Since the XX ground state is optically active, we deal with relatively
short XX lifetimes even at very low temperatures.
Such a low-temperature spectrum is plotted in
Fig.~\ref{fig:XandXX-hitemp}(lower left panel) with black bars, while
the black continuous line is the emission envelope accounting for a model
inhomogeneous broadening.
As the temperature is increased, excited XX states will become
populated.
This, however, does not have to mean that the emission peak will shift
towards higher energies, as was the case for the exciton.
Indeed, the XX states recombine to excited X states, while the X
states can only recombine to the vacuum.
Depending on the  oscillator strengths appropriate for each of the
possible XX-X transitions we may deal with additional higher-energy
peaks (when excited XX states recombine to low-lying X states) or
low-energy maxima (when the final X states lie high in energy).
Figure~\ref{fig:XandXX-hitemp}(lower right panel) demonstrates that in
the studied system we deal with the latter case.
If we also account for inhomogeneous broadening (red continuous line),
we find that the overall emission maximum moves towards {\em lower}
energies, i.e., exhibits a trend opposite to that of the exciton.

Let us now compare the two spectra.
In the $4$ K case (the two left panels), due to the interplay of
Coulomb elements, we 
predict the XX emission peak at a slightly higher energy than the X
maximum, i.e., we find that the XX becomes unbound.
However, as the temperature is raised to $300$ K (the two right
panels), the opposite shifts in the X and XX spectra lead to a
rearrangement of the order of the peaks, so that the XX maximum is
found below the X. 

In the last element of our analysis we study  the relative
positions of the X and XX emission maxima as a function of the NC
size.
In Fig.~\ref{fig:unbindingsize} we show several characteristic
quantities measuring the position of the XX peak relative to the X
peak, i.e., the XX binding energy.
Black squares show the difference $E_{XX0}-2E_{X0}$ computed earlier 
in the single-configuration approach neglecting the electron-hole
exchange.
As can be seen, this difference is positive (i.e., the XX is
unbound\cite{efros_rodina_ssc89}) 
for all NC sizes studied.
Next we include the electron-hole exchange effects and correlations by
computing the XX binding energy with the basis of $M_e=8$ and $M_h=28$
single-particle states.
In this case we present two sets of results, shown with filled symbols.
The red (blue) circles show the difference between the XX
lowest-energy peak and that of the dark (bright) exciton; the
splitting between these two sets of data is due to the electron-hole
exchange.
We find that if the position of the XX peak is measured relative to
the dark exciton, the XX becomes unbound for NC diameters smaller than
$4$ nm.
On the other hand, if the bright exciton is considered, the XX is
always bound, in contrast to the single-configuration calculation.
However, as demonstrated earlier, the basis set taken in this
calculation is not sufficient to achieve convergence of the X and XX
energies.
To eliminate this systematic error, we extrapolate the X and XX
energies presented in Fig.~\ref{fig:x-baseladders}
and~\ref{fig:XX-baseladders}, respectively, to the infinite hole basis.
The XX binding energy computed in this limit relative to the bright X
maximum is shown in Fig.~\ref{fig:unbindingsize} with empty symbols.
We find that the XX is unbound for NC diameters smaller than $4$ nm,
and bound for larger NC diameters.
We compare our results to those obtained by Sewall {\em et al.}
(Ref.~\onlinecite{sewall_franceschetti_prb09}) in a finite electron
and hole basis set.
The XX binding energy obtained in this empirical pseudopotential
calculation is denoted by the blue triangle for the NC diameter of
$3.8$ nm.
In agreement with our finite-basis calculation, Sewall {\em et al.}
predict a bound XX.
Note, however, that after extrapolation to the infinite basis set 
the XX becomes unbound, which demonstrates the need for systematic
convergence study.

\section{Conclusions}

In conclusion, we have analyzed the electronic and optical
properties of an exciton and a bi-exciton confined in a single,
spherical CdSe nanocrystal. 
Using the atomistic tight-binding approach we have calculated the
single-particle spectra  and found that the lowest energy hole states
form a shell consisting  of four states separated from the
rest of levels by a gap. 
The bi-exciton state was computed using configuration interaction
techniques and found to be a strongly correlated state consisting of a
two electron singlet in the $s$-shell of the conduction band and a
strongly correlated state of two holes distributed on the degenerate
hole shell, resulting in a fine structure of bi-exciton energy
levels. 
The fine structure is also present in the exciton spectrum, however it
is due to the electron-hole exchange interaction. 
The bi-exciton fine structure becomes apparent in the absorption of
the second exciton into the nanocrystal.
We find that if the initial state exciton is prepared in the bright
configuration, the maximum indicating the absorption of the second
pair is found at the same energy as the emission peak from XX.
However, if we prepare the exciton in the dark state, the absorption
takes place to higher XX states in the quasi-degenerate manifold.
As for the emission spectra, we found that at a low temperature the
bi-exciton emission peak corresponds to an energy slightly higher than
the energy of the excitonic ground, dark state, i.e., the XX is unbound.
However, at higher temperatures the exciton emits from the excited
bright state, so that the inhomogeneously broadened X emission peak
moves to higher energies.
On the other hand, thermal population of higher XX levels leads to
emission to excited final X states, moving the broadened XX emission
peak to lower energies, even below the high-temperature emission maximum.
A similar transition in the character of XX can be achieved by
changing the diameter of the nanocrystal: for diameters of up to $4$
nm the XX is unbound, while for larger NCs the XX becomes bound.
Due to the complicated nature of the spectrum of the valence hole we
find that in all elements of our analysis the correlations play a
crucial role and that any qualitative conclusions as to the electronic
and optical properties of X and XX can be drawn only after a careful
convergence analysis.

Future work will focus on improving several aspects of QNANO which 
at present received only a model treatment.
We plan to improve the description of screening and carry out a
computation of the distance-dependent dielectric
function.\cite{wang_califano_prl03,franceschetti_fu_prb99,moreels_allan_prb10,ogut_burdick_prl03,delerue_lannoo_prb03}
We are going to develop a more realistic model of the surface
passivation accounting for the presence of ligands.
The more microscopic description of the surface is also needed for a
proper description of the NC shape.
We plan to develop a hybrid tb-DFT QNANO package in order to
account for a realistic surface reconstruction and faceting.

\section*{Acknowledgment}
\begin{figure}[ht]
\includegraphics[width=0.45\textwidth]{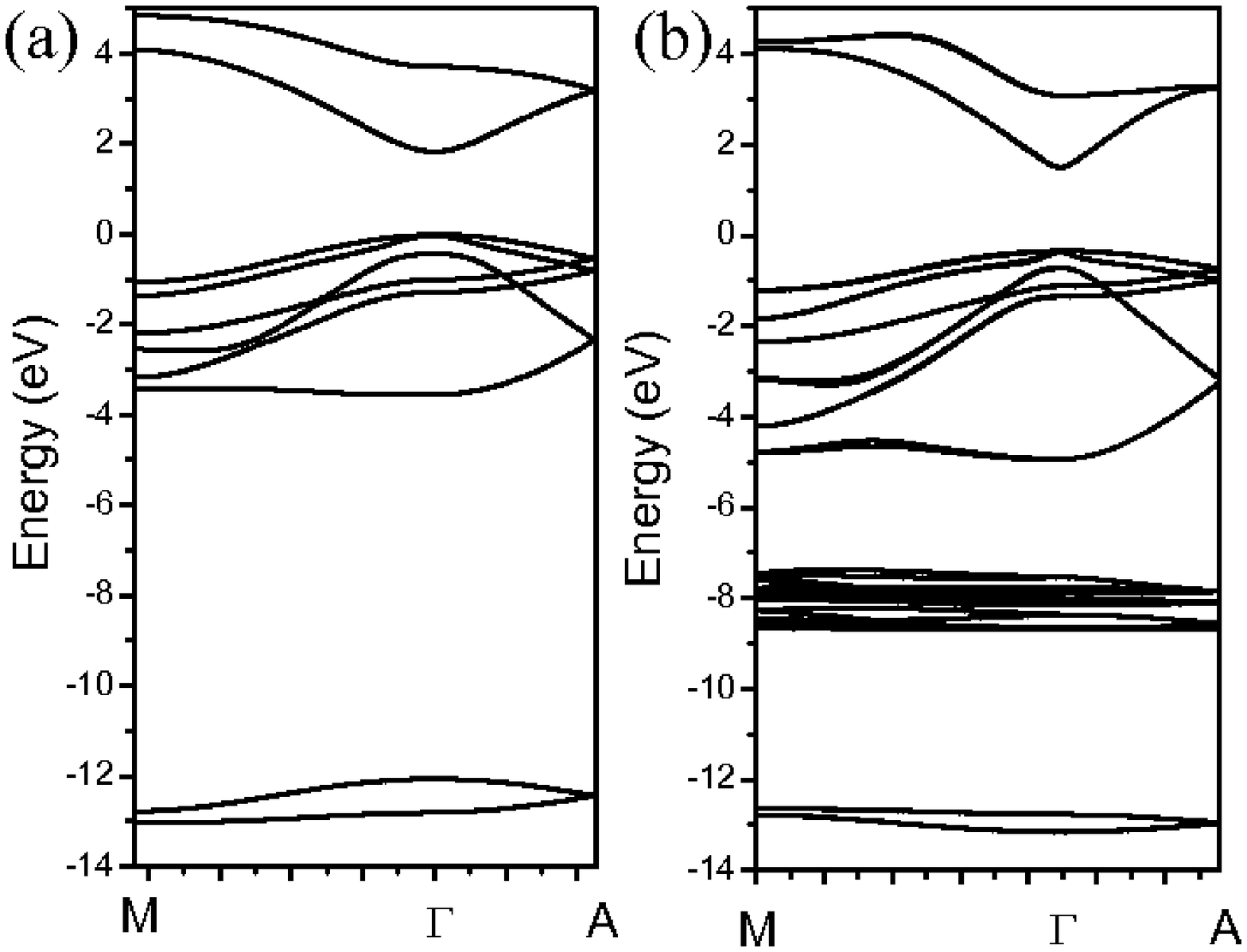}
\caption{Band structure of CdSe computed using the 20-band
  tight-binding model of this work (a) and the DFT approach with a
  rigid shift applied to the conduction band (b)}
\label{fig1}
\end{figure}

\begin{figure}[ht]
\includegraphics[width=0.45\textwidth]{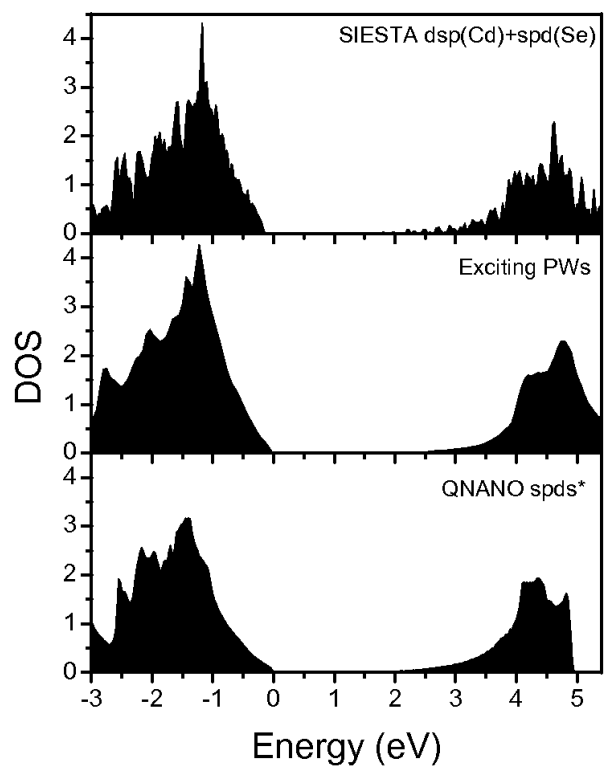}
\caption{Bulk density of states computed using the 
DFT procedure of SIESTA (top panel),plane-wave approach
of the package ''Exciting'' (middle panel),  
and our tight-binding model (bottom panel).}
\label{fig:bulk_dos}
\end{figure}

\begin{figure}[ht]
\includegraphics[width=0.45\textwidth]{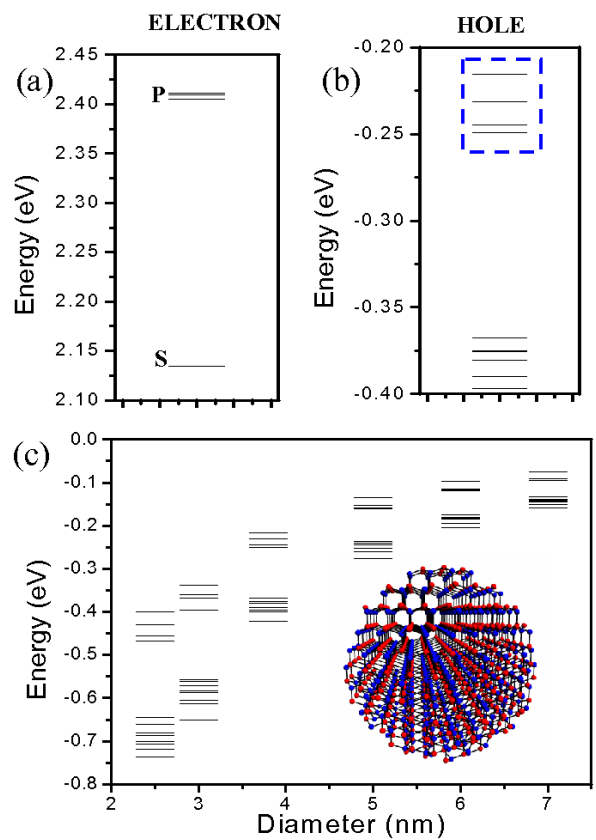}
\caption{(Color online)
Energies of single-particle states of an electron (a) and a hole (b)
in a CdSe nanocrystal of $3.8$ nm diameter. (c) Energies of the hole
states as a function of the diameter of the nanocrystal; the
characteristic gap separating the four lowest hole states from the
rest of the spectrum is visible for all nanocrystal sizes. Inset shows
an atomistic picture of the $3.8$ nm nanocrystal.}
\label{fig2}
\end{figure}

\begin{figure}[ht]
\includegraphics[width=0.45\textwidth]{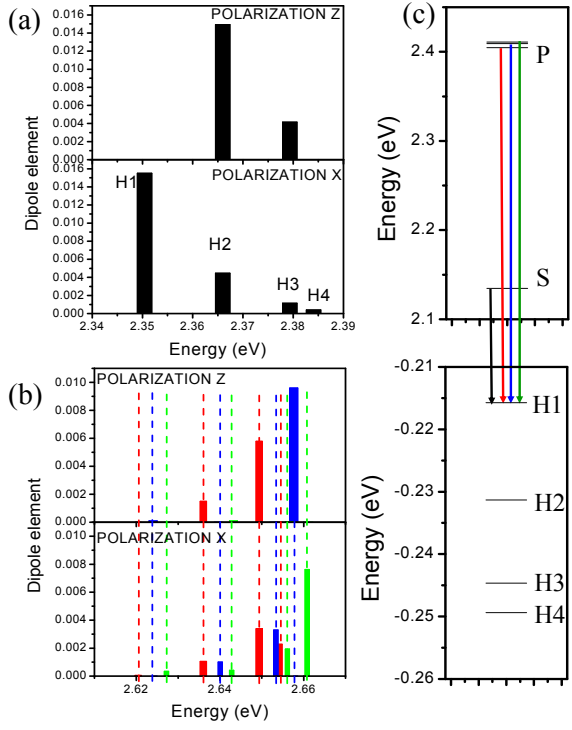}
\caption{(Color online)
Joint optical density of states characterizing the oscillator strength
between each of the four lowest hole states H1-H4 with the electron
$s$-shell (a) and $p$-shell (b) states. The color coding is explained in
part (c).}
\label{fig4}
\end{figure}

\begin{figure}[ht]
\includegraphics[width=0.45\textwidth]{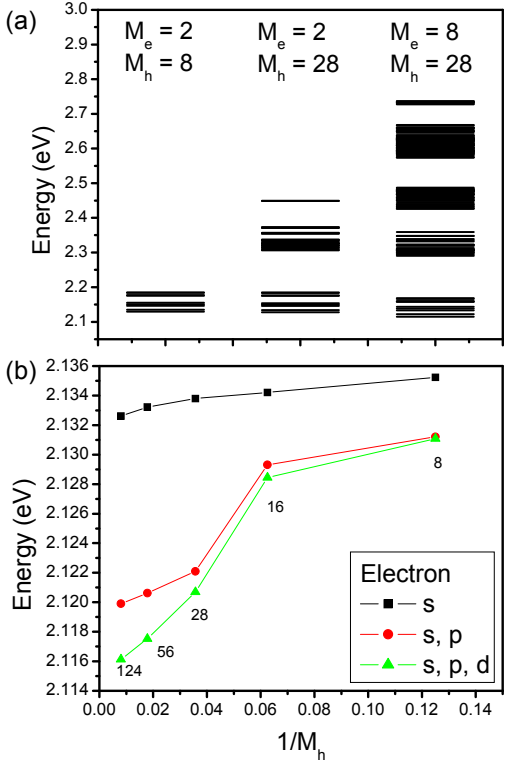}
\caption{ (Color online)
(a) Ground and excited energy levels of an exciton with
  increasing electron and hole single-particle basis size
for a nanocrystal with diameter of $3.8$ nm. 
(b) Ground-state energy of the exciton plotted as a function of
inverted hole basis size for $M_e=2$ (black squares), $M_e=8$ (red
circles), and $M_e=18$ (green triangles).}
\label{fig:x-baseladders}
\end{figure}

\begin{figure}[ht]
\includegraphics[width=0.45\textwidth]{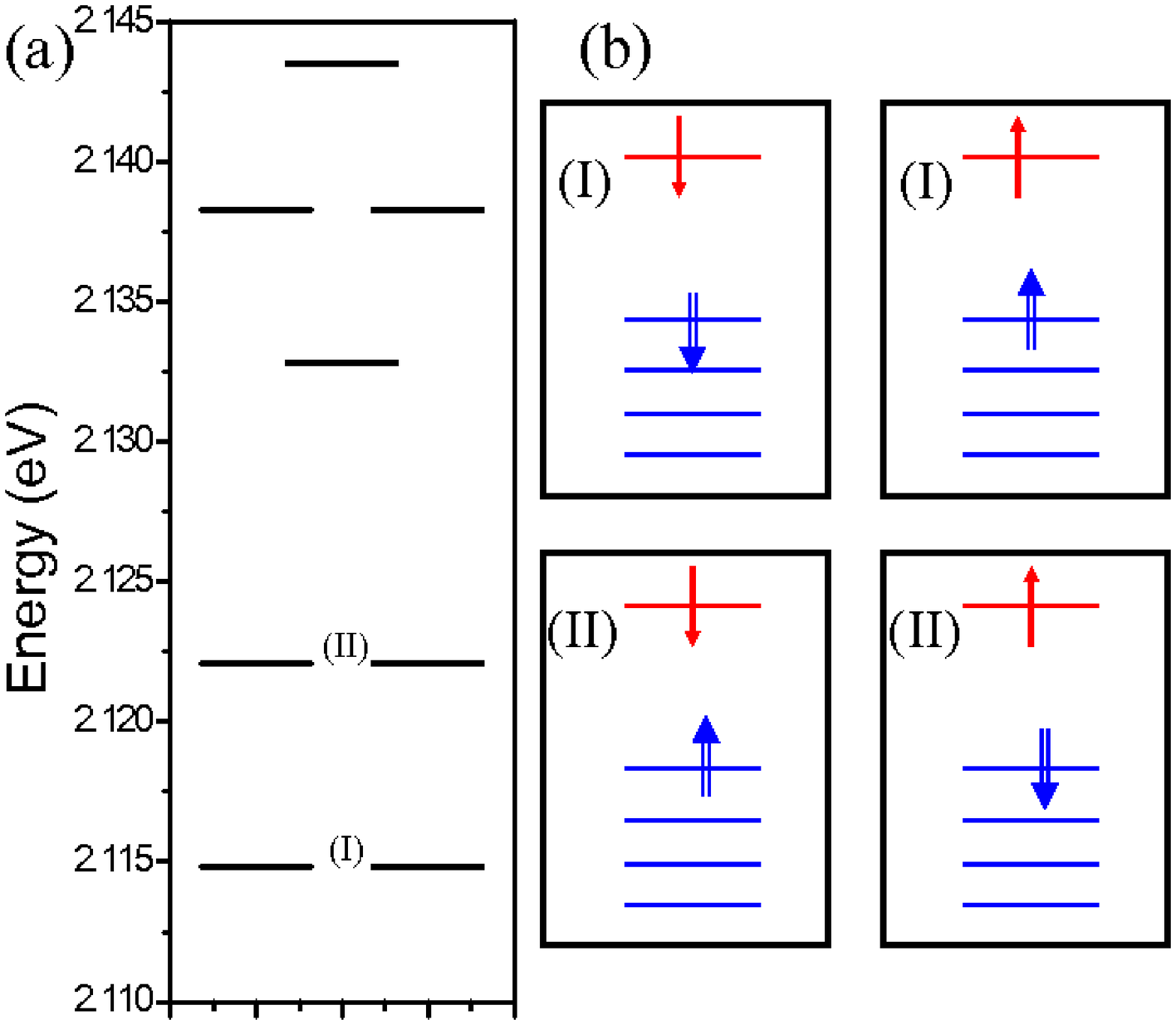}
\caption{(Color online)
(a) Ground and excited exciton energy levels computed with a basis of
$M_e=8$ electron and $M_h=28$ single-particle hole states
for a nanocrystal with diameter of $3.8$ nm. 
(b) Spectral content of the four lowest exciton states (see text for
analysis).}
\label{fig:X-converged-content}
\end{figure}

\begin{figure}[ht]
\includegraphics[width=0.45\textwidth]{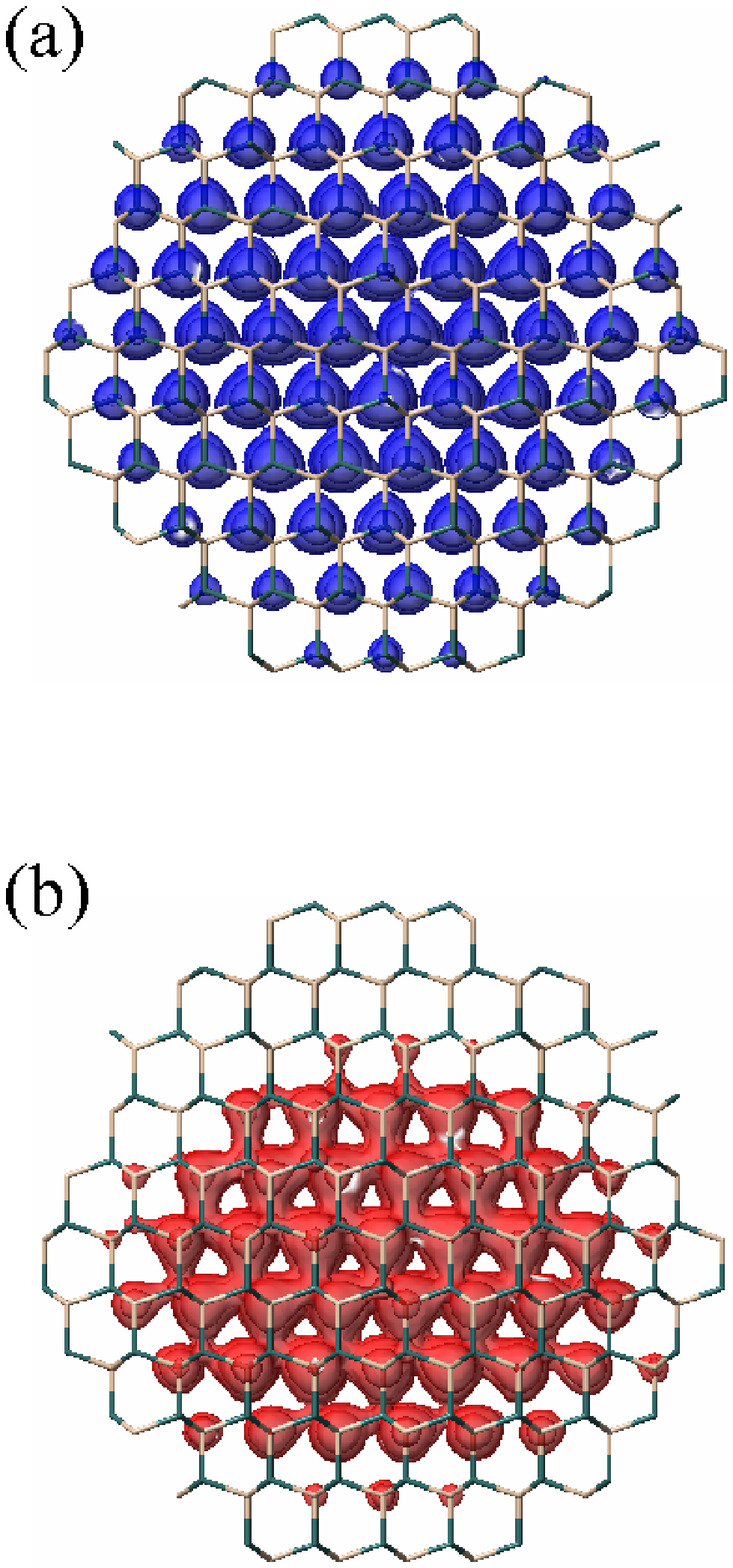}
\caption{(Color online)
Vertical cross-section of the
electron (a) and hole (b) ground state charge density in a nanocrystal
of $3.8$ nm diameter computed by the QNANO package.
}
\label{fig3}
\end{figure}

\begin{figure}[ht]
\includegraphics[width=0.45\textwidth]{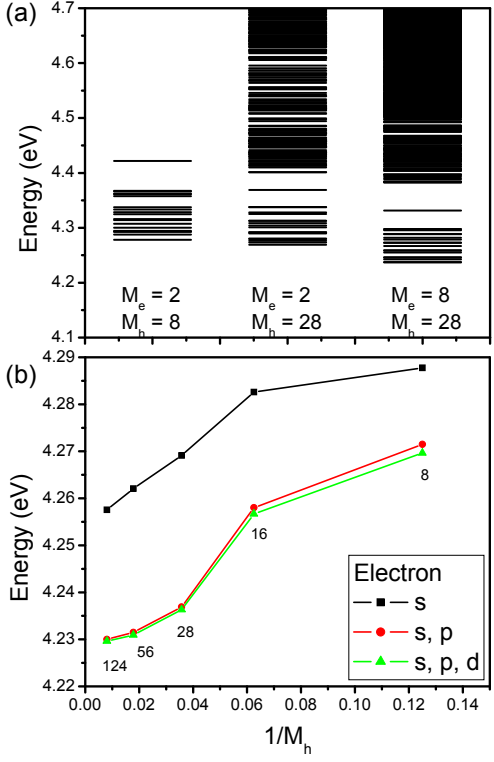}
\caption{ (Color online)
(a) Ground and excited energy levels of a bi-exciton with
  increasing electron and hole single-particle basis size
for a nanocrystal with diameter of $3.8$ nm. 
(b) Energy of the bi-exciton singlet-singlet state plotted as a function of
inverted hole basis size for $M_e=2$ (black squares), $M_e=8$ (red
circles), and $M_e=18$ (green triangles).}
\label{fig:XX-baseladders}
\end{figure}

\begin{figure}[ht]
\includegraphics[width=0.45\textwidth]{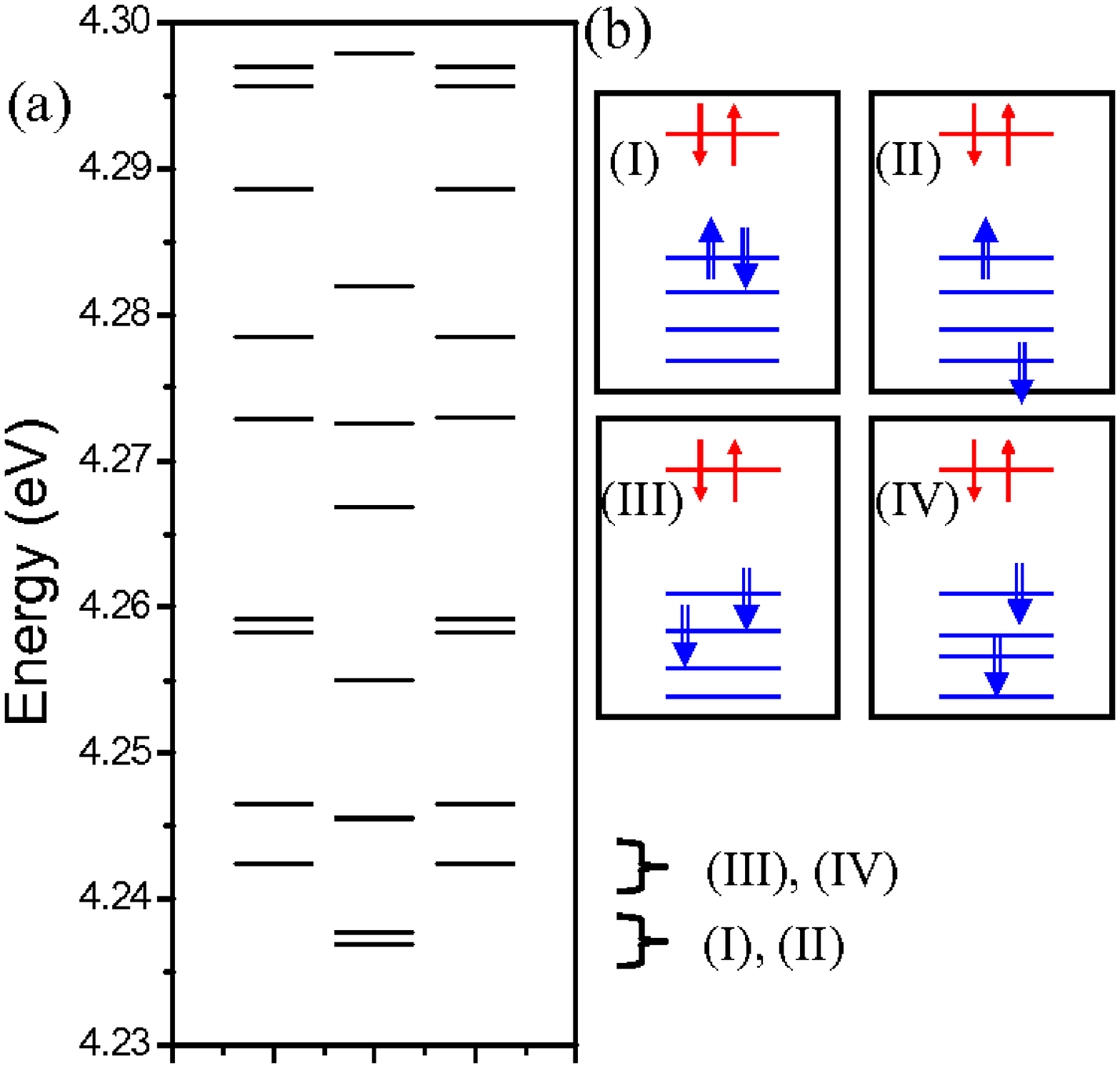}
\caption{ (Color online)
(a) Ground and excited bi-exciton energy levels computed
  with a basis of $M_e=8$ electron and $M_h=28$ single-particle hole
  states for a nanocrystal with diameter of $3.8$ nm.
(b) Spectral content of the four lowest bi-exciton states (see text for
analysis).}
\label{fig:XX-converged-content}
\end{figure}

\begin{figure}[ht]
\includegraphics[width=0.45\textwidth]{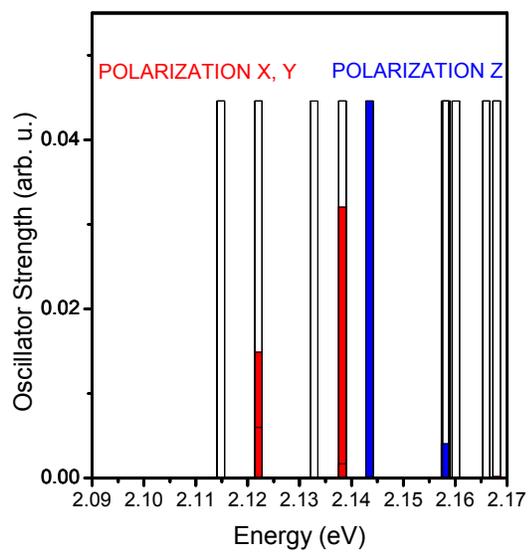}
\caption{(Color online)
Absorption spectrum of the exciton in polarization $x$ and
  $y$ (red bars), and $z$ (blue bars). 
Black bars denote positions of absorption maxima, whilst the height of
the color bars denotes the corresponding oscillator strength.} 
\label{fig:Xabsorption}
\end{figure}

\begin{figure}[ht]
\includegraphics[width=0.45\textwidth]{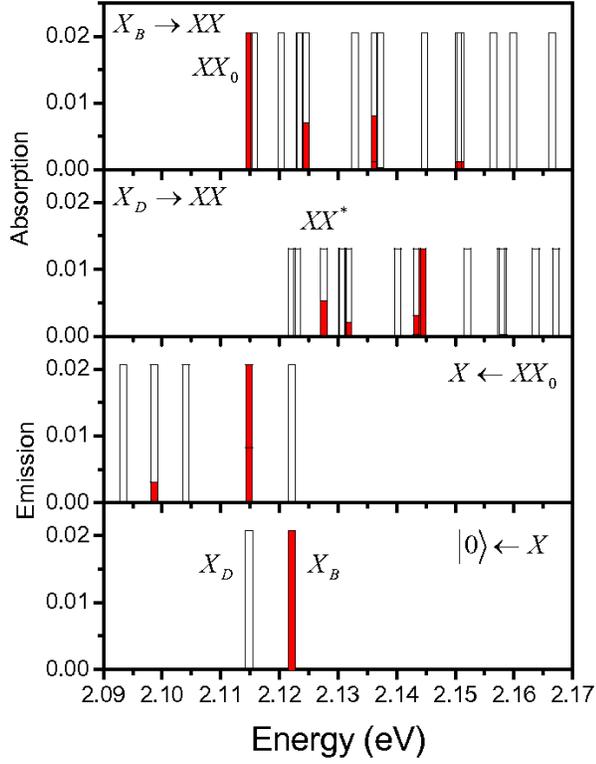}
\caption{(Color online)
Optical spectra of the exciton and bi-exciton for a nanocrystal with
diameter of $3.8$ nm. 
Upper panel shows the absorption spectrum of the second exciton
assuming that the first exciton is prepared in the bright excited
state. 
Second panel from the top shows the same assuming that the first exciton is
prepared in the dark ground state. 
Third panel shows the emission spectra from the ground bi-exciton to the
ground and excited exciton  states, whilst the bottom panel shows the
exciton emission spectrum. 
In all panels, black bars denote positions of absorption or emission
maxima, whilst the height of the red bars denotes the corresponding
oscillator strength.} 
\label{fig:XXoptics}
\end{figure}

\begin{figure}[ht]
\includegraphics[width=0.45\textwidth]{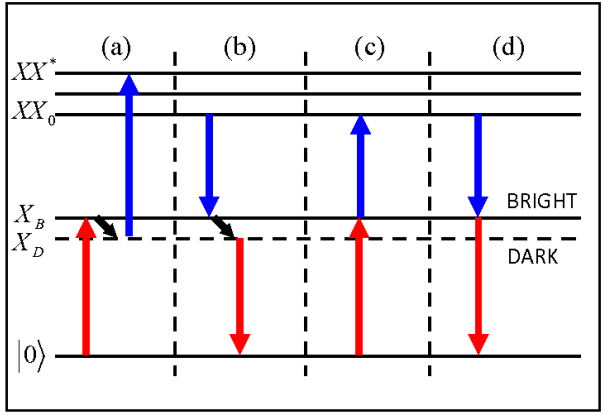}
\caption{ (Color online)
Schematic view of the exciton and biexciton absorptive and
  emissive processes assuming fast (a and b) and slow (c and d)
  relaxation from the excited, bright to the ground dark state of the
  exciton. Panels (a) and (c) show absorption processes, whilst panels
  (b) and (d) show emission processes.
}
\label{fig:abs_emis_processes}
\end{figure}

\begin{figure}[ht]
\includegraphics[width=0.45\textwidth]{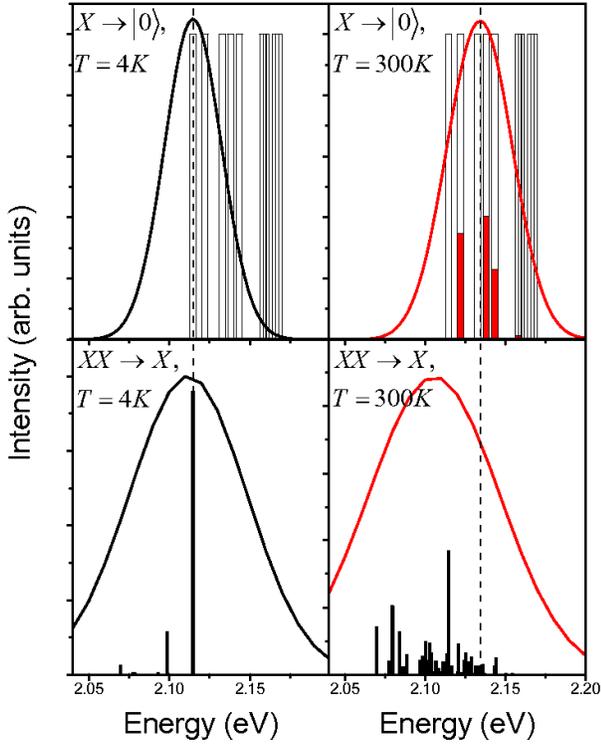}
\caption{(Color online)
Exciton (top panels) and bi-exciton (bottom panels) emission spectra at
low (left panels) and high temperatures (right panels).
Black (red) curves show emission spectra at low (high) temperature
accounting for model inhomogeneous broadening of $50$ meV.
Void bars denote positions of the respective maxima, whilst the height
of the solid bars represents the respective oscillator strength
multiplied by thermal population of levels.}
\label{fig:XandXX-hitemp}
\end{figure}

\begin{figure}[ht]
\includegraphics[width=0.45\textwidth]{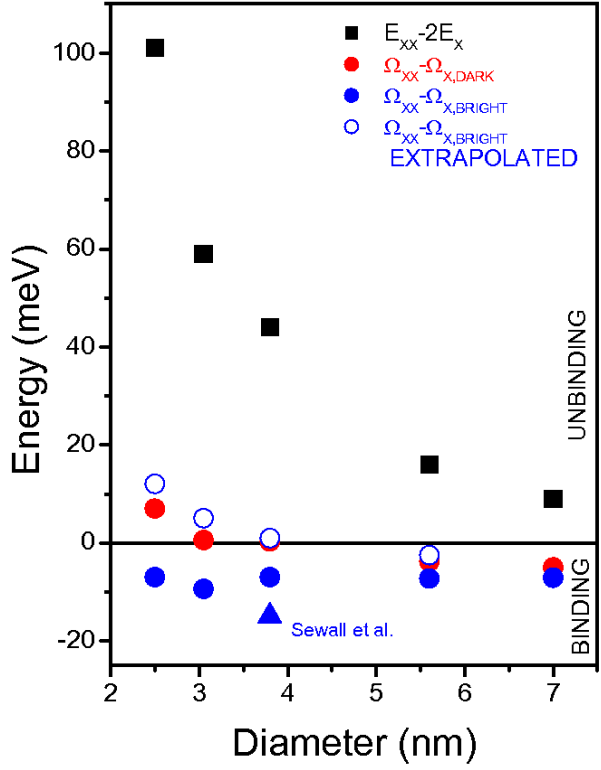}
\caption{(Color online)
Relative position of exciton and bi-exciton emission maxima at zero
temperature as a function of the nanocrystal size.
Black symbols show the bi-exciton binding energy calculated in a
single-configuration approach neglecting the electron-hole exchange
and correlations.
Full circles  denote the bi-exciton binding energy computed in the
basis of $M_e=8$ and $M_h=28$ states; the blue (red) symbols 
are computed in reference to the exciton bright (dark) emission peak.
Empty symbols show the bi-exciton emission peak relative to
the exciton bright transition obtained by extrapolation to infinite
electron and hole basis.
The triangle shows the XX binding energy obtained in empirical
pseudopotential calculation of Sewall et al. (Ref.~\onlinecite{sewall_franceschetti_prb09}).}
\label{fig:unbindingsize}
\end{figure}

The authors acknowledge discussions with G. Scholes, Kui Yu,
A. Stolow, P. Kambhampati, A. Efros, M. Zielinski, and thank
E. Kadantsev for providing the results of DFT calculations with the
EXCITING/elk software (Figs.~\ref{fig1},~\ref{fig:bulk_dos}). 
Funding from the NRC-NSERC-BDC Nanotechnology Project and CIFAR
is gratefully acknowledged.

\newpage

\end{document}